\documentclass[prb,twocolumn,amsmath,amssymb,nopacs,nofootinbib,floatfix]{revtex4}

\usepackage{bm}
\usepackage{graphicx}
\usepackage[normalem]{ulem}
\usepackage{hyperref}

\makeatletter

\begin{document}

\title{Non-compact lattice Higgs model with Abelian discrete gauge groups: \\ 
phase diagram and gauge symmetry enlargement}

\author{Claudio Bonati}
\affiliation{Dipartimento di Fisica dell'Universit\`a di Pisa
        and INFN, Largo Pontecorvo 3, I-56127 Pisa, Italy}

\author{Niccol\`o Francini}
\affiliation{Dipartimento di Fisica dell'Universit\`a di Pisa
        and INFN, Largo Pontecorvo 3, I-56127 Pisa, Italy}

\date{\today}

\begin{abstract}
We study the phase diagram and phase transitions of the three dimensional
multicomponent lattice Higgs model with non-compact Abelian discrete groups.
The model with non-compact U(1) gauge group is known to undergo, for a
sufficiently large number of scalar fields $N$, a continuous transition
associated to the charged fixed point of the continuous Abelian Higgs field
theory. We show that in the model with gauge group $\mathbb{Z}_q^{(nc)}\equiv 2\pi\mathbb{Z}/q$ only
critical transitions in the orthogonal universality classes are present for
small values of $N$, while a symmetry enlargement to the continuous Abelian
Higgs universality class happens when $q\ge 5$ and $N$ is large enough.
\end{abstract}

\maketitle


\section{Introduction}

Global symmetries and their spontaneous breaking play an essential role in
condensed matter physics, where they are used to classify phases of matter and
phase transitions since late 1930s~\cite{LL5, A_basic}. More recently, global
symmetries played a pivotal role in the modern theory of critical phenomena and
renormalization group~\cite{Wilson:1973jj, Z_quant, Pelissetto:2000ek}, which
clarified the relation between continuous phase transitions, 
symmetry breaking and quantum field theories.

In this framework universality classes are associated to the symmetry breaking
pattern of the effective Hamiltonian at the fixed point (FP) of the
renormalization group (RG) flow, and not to that of the microscopic Hamiltonian.
This allows for the existence of symmetry enlargements, associated to the emergence of new
symmetries at the critical point. This happens when the symmetry group
of the FP effective Hamiltonian is larger than that of the microscopic
Hamiltonian.  A simple model displaying symmetry enlargement is the three
dimensional $q$-state clock model, whose symmetry group is $\mathbb{Z}_q
\subsetneq O(2)$ but whose critical point is in the $O(2)$ universality
class for~\cite{Hove:2003qnt} $q\ge 5$, see also Refs.~\cite{Hasenbusch:2019jkj,
Hasenbusch:2020pwj} for similar cases.

Despite having been originally introduced in high energy
physics~\cite{W_the_q}, gauge theories are by now known to be ubiquitous also
in condensed matter physics~\cite{Fradkin_book, Moessner_book, Sachdev:2018ddg}, not to
mention the condensed matter side of high energy physics (see e.g.
Refs.~\cite{Pisarski:1983ms, Rajagopal:2000wf}). It is thus fundamental to
understand the critical behavior of models which are characterized both by
global and local symmetries. Multicomponent scalar models~\cite{Fradkin:1978dv}
appear to be ideal candidates for this purpose: their critical properties can
in some cases be determined or at least guessed by analytical methods, moreover
they are quite easy to study by numerical simulations.  The aim of this paper
is to investigate by Monte Carlo simulations a multicomponent lattice scalar
model with discrete Abelian gauge group, to understand if symmetry enlargement
is possible at a second order phase transition in which both gauge and matter
degrees of freedom are critical.

To put this statement in context it is convenient to recall some
facts about critical phenomena in gauge theories. Indeed three different
scenarios can be realized at the critical point of a model displaying both
local symmetries, constraining the form of the interactions, and global
symmetries, associated to the transformation properties of the matter fields. 

In the first scenario gauge fields simply act as spectators at the transition,
without developing long range correlations. In this case the only role of the
local invariance is that of preventing some modes (the non gauge invariant
ones) from acquiring non-vanishing expectation values. The critical behavior
can be modelled by using a local gauge invariant order parameter, and
everything goes on exactly as if no gauge symmetry were present. This happens
in the multicomponent compact lattice Abelian Higgs model~\cite{Pelissetto:2019zvh, 
Pelissetto:2019iic, Pelissetto:2019thf}, in models
with compact discrete Abelian symmetry~\cite{Bonati:2021thy, Bonati:2022sgs}
and in most of the non-Abelian models studied so far~\cite{Bonati:2019zrt,
Bonati:2020elf, Bonati:2021rzx, Bonati:2021tvg}.

The second scenario is the dual of the first one: matter fields remain
non-critical, while gauge modes develop long range order.  Just like the
transitions in pure gauge models~\cite{Wegner:1971app, Savit:1979ny,
Borisenko:2013xna}, transitions in this class are characterized by the absence
of a local order parameter, and they are thus called topological transitions.
Examples of this behavior are found in the multicomponent non-compact lattice
Abelian Higgs model~\cite{MV-08, KMPST-08, Bonati:2020jlm}, in the
multicomponent compact lattice Abelian Higgs model with charge $Q\ge
2$ (see Refs.~\cite{Bonati:2020ssr, Bonati:2022oez}), and also in some non-Abelian
models~\cite{Sachdev:2018nbk, Scammell:2019erm, Bonati:2021tvg}.

Finally, the third scenario is the one in which both the gauge and the matter
fields becomes critical at the transition. When this happens, a local gauge
invariant order parameter exists, but an effective field theory description of
the critical behavior requires to explicitly use
 both matter and gauge fields in the effective
Hamiltonian.  It should be clear that
transitions of this class are the most peculiar ones, and this is the case that
is usually refereed to as ``beyond the Landau-Ginzburg-Wilson paradigm''
\cite{SBSVF-04}.  At present we however know only few classical lattice models
exhibiting this type of critical transitions: compelling evidence has been
found for the multicomponent non-compact lattice Abelian Higgs
model~\cite{Bonati:2020jlm} and the multicomponent compact lattice Abelian
Higgs model with charge $Q\ge 2$ (see Refs.~\cite{Bonati:2020ssr, Bonati:2022oez}), while
for non-Abelian gauge models we only have hints of this type of
behavior~\cite{Bonati:2021tvg, Bonati:2021rzx}.

Let us now go back to symmetry enlargements in gauge models. When gauge fields
are non-critical, symmetry enlargements are known to happen, with examples of
continuous global O(2) symmetry emerging from discrete global $\mathbb{Z}_q$
symmetries reported e.g. in Refs.~\cite{Bonati:2021thy, Bonati:2022sgs}.
Symmetry enlargements have also been observed in pure gauge theories (see e.g.
Ref.~\cite{Borisenko:2013xna}), thus it seems reasonable to guess the same
phenomenon to be present also in the more general case of the second scenario
above. The case in which both gauge and matter fields are critical is the less
studied one, and the question of the existence of symmetry enlargement is still
open\footnote{A different kind of emergent symmetry was observed in
Ref.~\cite{Bonati:2022yqh}, in which two dimensional models with related numbers of
scalar flavors and colors turned out to have the same continuum limit.}.

To answer this question we study a variant of the non-compact lattice Abelian
Higgs model with $N$ scalar fields, and specifically the variant in which the
gauge field is restricted to the non-compact proper subgroup
$\mathbb{Z}_q^{(nc)}\equiv 2\pi\mathbb{Z}/q$ of U(1)$^{(nc)}=\mathbb{R}$ (we
denote by a superscript $nc$ the non-compact groups, in order to avoid
confusion with the compact ones).  The lattice model with gauge group
U(1)$^{(nc)}$ is indeed known to exhibit, for $N\gtrsim 10$, critical
transitions governed by the charged (i.e. with non-vanishing gauge coupling) FP
of the continuous Abelian Higgs field theory~\cite{Bonati:2020jlm}, thus
realizing the third scenario described above.

It is natural to expect the phase diagram of the $\mathbb{Z}_{q}^{(nc)}$ model
to approach that of the model with gauge group U(1)$^{(nc)}$ in the limit
$q\to\infty$.  Our main aim is to understand if a finite value $q^*$ exists
such that for $q\ge q^*$ the $\mathbb{Z}_{q}^{(nc)}$ model displays transitions
of the continuous Abelian Higgs universality class, as the U(1)$^{(nc)}$ model.
We thus investigate the phase diagram and phase transitions of the
$\mathbb{Z}_{q}^{(nc)}$ model for several values of $q$, and for $N$ values
below ($N=2$) and above ($N=25$) the threshold for the
appearance of the charged FP in the continuous Abelian Higgs model. 

A similar strategy has been very recently adopted in
Ref.~\cite{discretecompact}, where a $\mathbb{Z}_q$ deformation of the compact
U(1) lattice Abelian Higgs model with charge $Q=2$ (see Refs.~\cite{Bonati:2020ssr,
Bonati:2022oez}) was investigated.  By studying the region of the parameter
space where transitions of the continuous Abelian Higgs universality class
could emerge, the Authors found however only first order transitions for values
of $q$ up to $q=10$. 

The paper is organized as follows: in Sec.~\ref{sec:model} we summarize 
the main features of the phase diagram of the lattice U(1)$^{(nc)}$ model, then
we introduce the lattice $\mathbb{Z}^{(nc)}_{q}$ model and provide
arguments to delineate its phase diagram. In Sec.\ref{sec:num} we define the
observables that are used in the Monte Carlo simulations and we present the
numerical results obtained, discussing separately the case in which only matter
field are critical ($N=2$) and the case in which both gauge and matter
fields develop critical correlations ($N=25$).  Finally, in
Sec.~\ref{sec:concl} we draw our conclusions and discuss open problems to be
further investigated.

\section{The lattice model}
\label{sec:model}

\subsection{The U(1)$^{(nc)}$ lattice model}
\label{sec:u1}

The lattice Hamiltonian of the non-compact U(1)$^{(nc)}$ (equivalently
$\mathbb{R}$) Abelian Higgs model with $N$ scalar field flavors is
\begin{equation}
\begin{aligned}\label{eq:Hnc}
H&=H_z+H_g\ ,\\
H_z &= - J N \sum_{{\bm x}, \mu} 2\, {\rm Re}\,(e^{iA_{{\bm x},\mu}}\, 
\bar{\bm{z}}_{\bm x} \cdot {\bm z}_{{\bm x}+\hat{\mu}})\ ,\\
H_g&=\frac{\kappa}{2} \sum_{{\bm x},\mu>\nu} (\Delta_{\mu} A_{{\bm x},\nu} - 
\Delta_{\nu} A_{{\bm x},\mu})^2\ ,
\end{aligned}
\end{equation}
where $\bm x$ stands for a lattice point and $\mu,\nu=1,2,3$ denote the positive
directions along the axes.  In this expression ${\bm z}_{\bm x}$ represents a
$N$-component complex vector subject to the constraint $\bar{{\bm z}}_{\bm x}\cdot{\bm z}_{\bm x}=1$,
while the gauge field $A_{{\bm x},\mu}$ is a real number and the finite differences 
$\Delta_{\mu} A_{{\bm x},\nu}$ 
are defined by
\begin{equation}
\Delta_{\mu} A_{{\bm x},\nu} = A_{{\bm x}+\hat{\mu},\nu} - A_{{\bm x},\nu}\ .
\end{equation}
The partition function of the U(1)$^{(nc)}$ model is formally defined by the
expression (see later for a caveat)
\begin{equation}
Z=\sum_{\{{\bm z}_{\bm x}, A_{{\bm x},\mu}\}} e^{-\beta H}\ ,
\end{equation}
and in the following we will set $\beta=1$, which is equivalent to measure $J$ and $\kappa$ in units of $\beta$.

The Hamiltonian in Eq.~\eqref{eq:Hnc} is invariant under the global SU($N$) symmetry 
${\bm z}_{\bm x}\to M {\bm z}_{\bm x}$, with $M\in$ SU($N$), and under the local U(1) symmetry
\begin{equation}\label{eq:gaugeinv}
{\bm z}_{\bm x}\to e^{i\alpha_{\bm x}}{\bm z}_{\bm x}\ ,\  
A_{{\bm x},\mu}\to A_{{\bm x},\mu}+\alpha_{\bm x+\hat{\mu}}-\alpha_{\bm x}\ ,
\end{equation}
with $\alpha_{\bm x}\in\mathbb{R}$. The theory is also invariant under the
global transformation $A_{{\bm x},\mu}\to A_{{\bm x},\mu}+2\pi n_{\mu}$, where
$n_{\mu}$ is an integer depending only on the direction $\mu$, which is the
equivalent for this model of the center symmetry in compact lattice gauge
theories~\cite{McLerran:1981pb, Svetitsky:1982gs}.  This invariance makes the
partition function of the theory divergent, even after gauge fixing, on finite
lattices with periodic boundary conditions, and to make the theory well defined
on a finite lattice it was suggested~\cite{Bonati:2020jlm} to use the $C^*$
boundary conditions~\cite{Kronfeld:1990qu}
\begin{equation}\label{eq:cbc}
A_{{\bm r} + L_{\nu}\hat{\nu}, \mu} = - A_{{\bm r}, \mu}\ ,\quad
{\bm z}_{{\bm r} + L_{\nu}\hat{\nu}} = \bar{\bm z}_{\bm r}\ , 
\end{equation}
where $L_{\nu}$ is the lattice extent in the direction $\nu$.

\begin{figure}[t]
\includegraphics*[width=0.95\columnwidth]{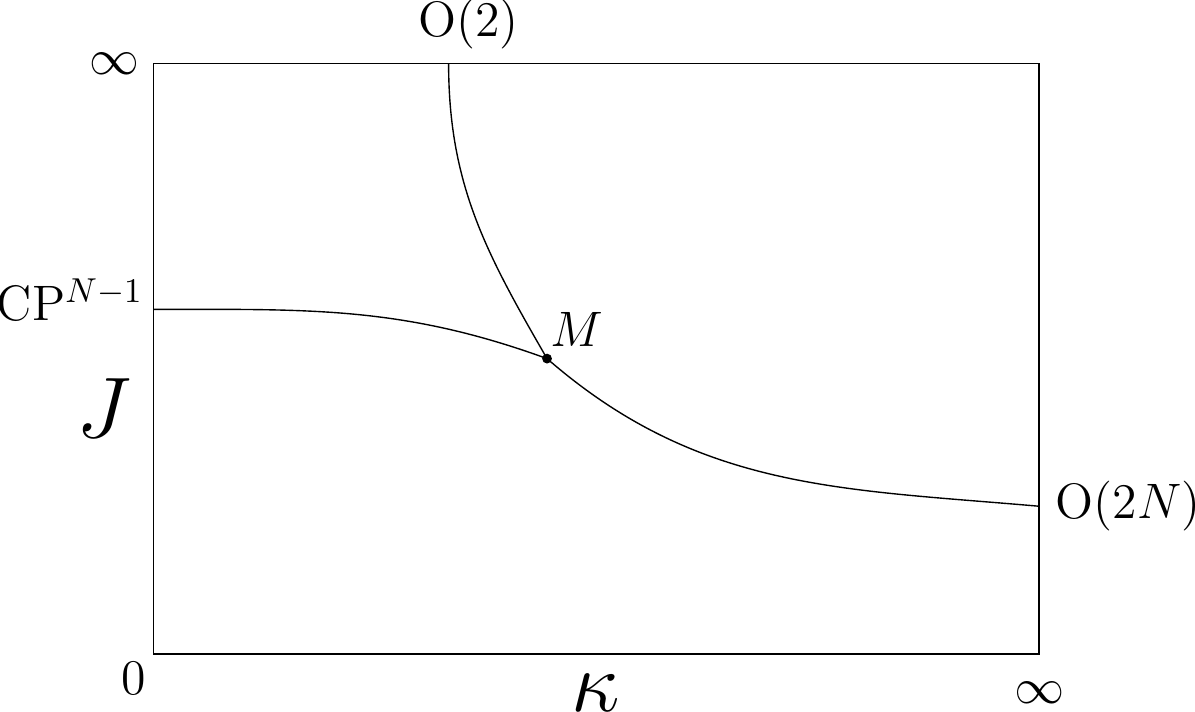}
  \caption{Qualitative sketch of the phase diagram of the lattice Abelian Higgs with gauge
  group U(1)$^{(nc)}$ and $N$ scalar flavors. 
  }
\label{fig:phd_u1}
\end{figure}

A sketch of the phase diagram of the lattice Abelian Higgs model with gauge
group U(1)$^{(nc)}$ is shown in Fig.~\ref{fig:phd_u1} (see Refs.~\cite{MV-08,
KMPST-08, Bonati:2020jlm}): three different thermodynamic phases exist, which
are separated by three transition lines and a multicritical point. To
understand the topology of the phase diagram it is convenient to look at 
the model for extremal values of the parameters, i.e. 0 or $\infty$
(see e.g. Ref.~\cite{Bonati:2020jlm} for more details). 

For $\kappa \to \infty$
the minimum of the Hamiltonian corresponds to 
\begin{equation} 
\Delta_{\mu} A_{{\bm x},\nu} - \Delta_{\nu} A_{{\bm x},\mu}=0\ , 
\end{equation} 
thus with a gauge transformation it is possible to set $A_{{\bm x},\mu}=0$ (in
the infinite volume limit). It is then simple to show that the model reduces to
the O($2N$) lattice model, which has a second order phase transition as a
function of $J$ for any $N$. For $\kappa=0$ we instead obtain the gauged form
of the lattice CP$^{N-1}$ model~\cite{Pelissetto:2019zvh,
Pelissetto:2019iic,Pelissetto:2019thf}, which displays as a function of $J$ a
second order transition of the O(3) universality class for $N=2$, and a first
order phase transitions for $N>2$. This transition is associated to the
spontaneous breaking of the global SU($N$) symmetry of the model, and the order
parameter is the gauge invariant bilinear
\begin{equation}\label{eq:Q}
Q_{{\bm x}}^{ab} = \bar{z}_{\bm x}^a z_{\bm x}^b - \frac{1}{N}\delta^{ab}\ .
\end{equation}

For $J=0$ the model reduces to a system of non-interacting lattice photons and
no phase transition is encountered by varying $\kappa$. In the $J\to\infty$
limit it can be shown that the only configurations with non-vanishing weight
are those with $A_{{\bm x},\mu}=2\pi m_{{\bm x},\mu}$, where $m_{{\bm
x},\mu}\in\mathbb{Z}$. By performing a duality
transformation~\cite{Dasgupta:1981zz, Neuhaus:2002fp} it is then possible to
obtain the Villain discretization of the O(2) model. As a consequence, for
$J\to\infty$ the U(1)$^{(nc)}$ model undergoes for any $N$ a topological
transition of the O(2) universality class (with inverted high and low
temperature phases) at
\begin{equation}\label{eq:kcu1}
\kappa_c^{\mathrm{U(1)}} (J = \infty) = 0.076051(2)\ . 
\end{equation}
This value is obtained from $\beta_c = 3.00239(6)$
reported in Ref.~\cite{Neuhaus:2002fp} with the identification
$\kappa_c=\beta_c/(2\pi)^2$. 

The transitions emerging from the boundaries of the phase diagram merge at a
multicritical point and delimit three different phases. The phase in the upper
left corner of Fig.~\ref{fig:phd_u1} is characterized by broken SU($N$)
symmetry and long range gauge correlations (it is the ``low temperature'' phase
of the inverted O(2) transition), the phase in the lower part of the diagram is
instead characterized by unbroken SU($N$) symmetry and long range gauge
correlations. Finally in the upper right phase of Fig.~\ref{fig:phd_u1} the
SU($N$) symmetry is broken and gauge correlations are short range. 

Along the line of phase transitions connecting the multicritcal point $M$ with
the O($2N$) asymptotic point, both matter and gauge field correlators change
their long distance behavior. For small values of $N$, transitions on this line
are of the first order~\cite{MV-08,KMPST-08, Bonati:2020jlm}, while for $N\gtrsim
10$ they become continuous transitions, whose critical properties are
consistent with those expected at the charged FP of the continuous Abelian
Higgs model. Indeed the critical exponents estimated from numerical
simulations~\cite{Bonati:2020jlm}, are consistent with those computed in the continuous model in the
large $N$ limit~\cite{HLM-74, Irkhin:1996drp, Moshe:2003xn}. Also the number of
flavors required for the existence of a second order phase transition along
this line is consistent with analytical results, coming from a constrained
resummation of the $\epsilon$-expansion of Abelian Higgs field
theory~\cite{Ihrig:2019kfv}.

\subsection{The $\mathbb{Z}_q^{(nc)}$ lattice model}

Having summarized the results obtained for the lattice model with U(1)$^{(nc)}$
gauge group, we can now easily introduce the lattice model with reduced gauge
symmetry $\mathbb{Z}_q^{(nc)}\equiv 2\pi \mathbb{Z}/q$ and discuss its possible
phase diagram.

The Hamiltonian of the model with gauge invariance $\mathbb{Z}_q^{(nc)}$ is
once again Eq.~\eqref{eq:Hnc}, but now the field $A_{{\bm x},\mu}$ is
not represented by a generic real number, but it is constrained to be of the form
\begin{equation}\label{eq:Azq}
A_{{\bm x},\mu}=\frac{2\pi}{q}n_{{\bm x},\mu}\ ,\quad 
n_{{\bm x},\mu}\in\mathbb{Z}\ .
\end{equation}
The global SU($N$) symmetry of the U(1)$^{(nc)}$ model is a symmetry also of
the $\mathbb{Z}_q^{(nc)}$ model, and the corresponding order parameter is the
same $Q^{ab}$ introduced in Eq.~\eqref{eq:Q}. Also the global symmetry $A_{{\bm
x},\mu}\to A_{{\bm x},\mu}+2\pi m_{\mu}$, with $m_{\mu}\in\mathbb{Z}$, is still
present. The local invariance is obviously $\mathbb{Z}_q^{(nc)}$, i.e. the
Hamiltonian is invariant under the transformation in Eq.~\eqref{eq:gaugeinv}
where $\alpha_{\bm x}$ is an integer multiple of $2\pi/q$.  Finally, due to the
reduced gauge invariance, also the global symmetry
$\mathrm{U(1)}^{(nc)}/\mathbb{Z}_q^{(nc)}=\mathrm{U(1)}/\mathbb{Z}_q$ is now
present, and a gauge invariant order parameter for its breaking is
\begin{equation}
O^{i_1\cdots i_q}=z_{\bm x}^{(i_1)}\cdots z_{\bm x}^{(i_q)}\ ,
\end{equation}
where $i_k\in \{1,\ldots,N\}$ and $z_{\bm x}^{(i)}$ stands for the $i$-th
component of ${\bm z}_{\bm x}$.  Note that this order parameter transforms
nontrivially under the global SU($N$) symmetry, while $Q^{ab}$ is invariant
under the $\mathrm{U(1)}/\mathbb{Z}_q$ global symmetry. As a consequence the
$\mathrm{U(1)}/\mathbb{Z}_q$ symmetry can be spontaneously broken only in a
phase in which SU($N$) is also broken.

Let us now discuss the phase diagram of the $\mathbb{Z}_q^{(nc)}$ lattice
model. As for the case of the U(1)$^{(nc)}$ model, to understand the structure
of the phase diagram it is convenient to start analyzing the extreme cases. In
the limit $J\to\infty$ and in the limit $\kappa\to\infty$ the model is exactly
equivalent to the U(1)$^{(nc)}$ model discussed in Sec.~\ref{sec:u1}. We thus
expect for $J\to\infty$ an inverted O(2) topological transition with critical coupling 
(see Eq.~\eqref{eq:kcu1})
\begin{equation}\label{eq:ZqJinf}
\kappa_c^{\mathbb{Z}_q}(J=\infty)=\kappa_c^{U(1)}(J=\infty)=0.076051(2)\ ,
\end{equation}
while for $\kappa\to\infty$ we expect a transition in the O($2N$) universality
class.  For $J=0$ the $\mathbb{Z}_q^{(nc)}$ model is equivalent to the
$J\to\infty$ limit of the U(1)$^{(nc)}$ model, up to the rescaling $\kappa\to
\kappa/q^2$. We thus expect also in this case an inverted O(2) transition with critical coupling
\begin{equation}\label{eq:ZqJzero}
\kappa_c^{\mathbb{Z}_q}(J=0)=q^2\kappa_c^{U(1)}(J=\infty)=q^2\,0.076051(2)
\end{equation}
for all $N$ values. 

What happens for $\kappa=0$ is already nontrivial, but it is natural to expect
the presence of two transitions: one at $J_{c1}$ at which the global SU($N$)
symmetry gets spontaneously broken, and another one at a value of the coupling
$J_{c2}>J_{c1}$, at which also the $\mathrm{U(1)}/\mathbb{Z}_q$ symmetry gets
broken. A priori the two transitions could also happen at the same point,
however it seems reasonable to assume the phase diagram of the
$\mathbb{Z}_q^{(nc)}$ lattice model to converge to that of the U(1)$^{(nc)}$
model for $q\to\infty$. Since in the U(1)$^{(nc)}$ model only a single
transition (the SU($N$) breaking one) is present for $\kappa=0$, if follows that
$J_{c2}\to\infty$ when $q\to\infty$, thus $J_{c2}$ is generically strictly
larger than $J_{c1}$.

\begin{figure}[t]
\includegraphics*[width=0.95\columnwidth]{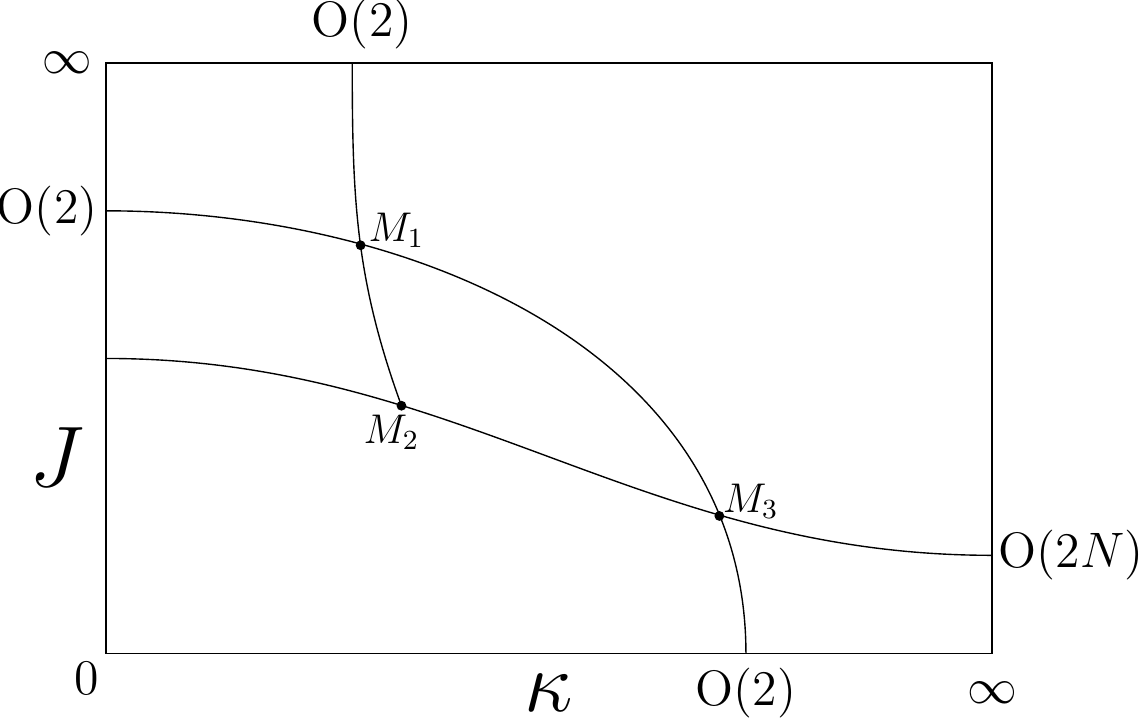}
  \caption{Qualitative sketch of the phase diagram of the lattice Abelian Higgs
  with gauge group $\mathbb{Z}_q^{(nc)}$ and $N$ scalar flavors. 
  }
\label{fig:phd_zq}
\end{figure}

The simplest topology of the phase diagram consistent with these boundary cases
is the one sketched in Fig.~\ref{fig:phd_zq}, in which six phases are separated
by several transition lines that intersects at three multicritical points. 

At the multicritical point denoted by $M_1$ in Fig.~\ref{fig:phd_zq} two O(2)
lines\footnote{To avoid complicating the discussion we assume all the lines to
correspond to continuous transitions, but obviously the presence of first order 
transitions can not be excluded.} cross each other, but the relevant degrees of
freedom of the two transitions are very different: the O(2) line starting from
$J=\infty$ is of topological nature, while the O(2) line starting from
$\kappa=0$ is associated to a global symmetry breaking. It thus seems natural
to guess the critical behaviors associated to these two lines to be decoupled
at the multicritical point $M_1$. If this holds true, $M_1M_2$ is a line of
O(2) topological transitions and $M_1M_3$ is a line of O(2) global symmetry
breaking transitions.

For small $\kappa$ values the gauge field always displays long range
correlations (since we are in the ``low temperature'' phase of both the inverted
O(2) topological transitions), and moving from small to large values of the
coupling $J$ we pass through two phase transitions, corresponding to the
spontaneous breaking of U($N$) and $\mathrm{U(1)}/\mathbb{Z}_q$ symmetries
respectively. For large values of $\kappa$, gauge field correlators are always
short range, and by increasing the coupling $J$ we meet a single transition, at
which both U($N$) and $\mathrm{U(1)}/\mathbb{Z}_q$ symmetries get spontaneously
broken. Since the lattice field strength $\Delta_{\mu} A_{{\bm x},\nu} -
\Delta_{\nu} A_{{\bm x},\mu}$ can only assume discrete values, for
$\kappa\to\infty$ the number of plaquettes on which the field strength is
nonvanishing is exponentially suppressed in $\kappa$. It is thus reasonable to
expect this transition line to be in the O($2N$) universality class.

The region of intermediate $\kappa$ values, roughly $0.076\lesssim
\kappa\lesssim q^2\,0.076$, is the most interesting one: for small values of the
$J$ coupling gauge field correlators are long range (we are in the ``low
temperature'' phase of the inverted O(2) transition departing from $J=0$) and no
symmetry breaking is present, but crossing the $M_2M_3$ line (see
Fig.~\ref{fig:phd_zq}) gauge field correlators become short range (we are in
the ``high temperature'' phase of the inverted O(2) transition $M_1M_2$) and
SU($N$) gets spontaneously broken.  By further increasing the coupling $J$ we
cross the $M_1M_3$ line and also the global $\mathrm{U(1)}/\mathbb{Z}_q$ gets
finally broken.

This phase diagram is consistent with that of the U(1)$^{(nc)}$ model: in the
large $q$ limit the multicritical points $M_1$ and $M_3$ move toward larger
and larger values of the couplings, while the multicritical point $M_2$ becomes
the multicritical point $M$ of the U(1)$^{(nc)}$ model. This phase diagram is
also very similar to the one discussed in Ref.~\cite{discretecompact}, where a
$\mathbb{Z}_q$ gauge version of the compact lattice Abelian Higgs model with charge
$Q=2$ was investigated. In this model the O(2) lines starting from $J=\infty$
and $J=0$ in Fig.~\ref{fig:phd_zq} becomes $\mathbb{Z}_Q$ and $\mathbb{Z}_q$
lines respectively, but apart from that the phase diagram looks the same.

To search for a symmetry enlargement when both gauge and matter degrees of
freedom are critical, the points to be investigated are the ones on the
$M_2M_3$ line, when $N$ is large enough that a transition of the continuous
Abelian Higgs universality class is present in the lattice  U(1)$^{(nc)}$
model.  In the next section we present the results of numerical simulations
performed along this line for $N=25$, which is large enough for the second
order transition of the continuous Abelian Higgs universality class to be
present in the U(1)$^{(nc)}$ model. We also report the results of some
simulations carried of for $N=2$ in the $\mathbb{Z}_q^{(nc)}$ lattice model,
with the purpose of checking whether continuous transitions of new universality
classes could appear in this case.

\section{Numerical results}
\label{sec:num}

Simulations have been performed on symmetric $L^3$ lattices using $C^*$
boundary conditions along all directions (see Eq.~\eqref{eq:cbc}). The gauge
field has been updated using the Metropolis algorithm, using $n_{{\bm
x},\mu}+1$ or $n_{{\bm x},\mu}-1$ as trial state with the same probability (see
Eq.~\eqref{eq:Azq}). Scalar fields have been updated using a combination of
Metropolis and overrelaxation updates, in the ratio of 1:5. A typical order of
magnitude of the statistics accumulated is of the order of $O(10^6)$
configurations for each data point, taken after 10
complete updates (Metropolis and overrelaxation) of the lattice, with the
autocorrelation time that was at most of the order of $O(10^3)$.

\subsection{Observables and finite size scaling}
\label{sec:obs}

The main observables used are the ones related to the spontaneous breaking of
the global SU($N$) symmetry, written by means of the gauge invariant hermitian order
parameter introduced in Eq.~\eqref{eq:Q}. 

From the two point function in momentum space $\tilde{G}(\bm p)$ of the
operator $Q_{\bm x}^{ab}$, defined by
\begin{equation}
\begin{aligned}
\tilde{Q}^{ab}(\bm p)&=\sum_{\bm x} Q_{\bm x}^{ab}e^{i{\bm p}\cdot{\bm x}}\ , \\
\tilde{G}(\bm p)&=\frac{1}{L^3}\mathrm{Re}\left\langle
\tilde{Q}^{ab}(\bm p)\tilde{Q}^{ab}(-\bm p)\right\rangle\ ,
\end{aligned}
\end{equation}
we can define the susceptibility 
\begin{equation}
\chi=\tilde{G}(\bm 0)
\end{equation}
and the second moment correlation length
\begin{equation}
\xi^2=\frac{1}{4\sin^2(\pi/L)}\frac{\tilde{G}(\bm 0)-\tilde{G}({\bm p}_m)}{\tilde{G}({\bm p}_m)}\ ,
\end{equation}
where ${\bm p}_m=(2\pi/L,0,0)$. Another useful quantity is the Binder cumulant 
\begin{equation}
U=\frac{\langle \mu_2^2\rangle}{\langle\mu_2\rangle^2}\ ,\quad \mu_2=\frac{1}{L^3}\mathrm{Re\,Tr}(\tilde{Q}(\bm 0)^2)\ ,
\end{equation}
which is a RG invariant quantity, just like $R_{\xi}=\xi/L$.

Renormalization group invariant quantities are particularly useful since their
finite size scaling (FSS) behavior at a second order phase transition is very simple. 
If we denote by $R$ a generic RG invariant quantity, its FSS is of the form
\begin{equation}
R=f_R(X)+L^{-\omega}g_R(X)\ ,
\end{equation}
where $f_R$ and $g_R$ are functions which are universal up to a rescaling of
their arguments, $\omega$ is related to the leading irrelevant RG exponent of the
transition and $X=(J-J_c)L^{1/\nu}$ or $X=(\kappa-\kappa_c)L^{1/\nu}$. Using
the two RG invariant quantities $R_{\xi}$ and $U$, it is possible write down a
FSS relation which is independent of any non-universal parameter and of the
critical exponents: 
\begin{equation}\label{eq:urxi}
U=F_U(R_{\xi})+O(L^{-\omega})\ .
\end{equation}
The function $F_U$ is universal, and depends only on some generic features of
the lattice, like the boundary conditions and the aspect ratio adopted. In the
following we will make extensive use of this relation to compare the results
obtained in the U(1)$^{(nc)}$ lattice model with those obtained in the
$\mathbb{Z}_q^{(nc)}$ lattice model. 

For comparison the FSS of the
susceptibility $\chi$ can be written in the form
\begin{equation}
\chi=L^{2-\eta_q}[f_{\chi}(R_{\xi})+O(L^{-\omega})]\ ,
\end{equation}
where we denoted by $\eta_q$ the anomalous dimension of the operator $Q^{ab}$.
In the following of the paper we mainly rely on the parameter-free scaling of
$U$ against $R_{\xi}$ to identify the universality class encountered, however
we have also checked that the scaling of $\chi$ against $R_{\xi}$ gives
consistent results. For the O(2N) transition at large $\kappa$, it can be shown
that $\eta_q$ is associated to the RG exponent $Y_2$ of the spin 2 operator of
Ref.~\cite{HV:aniso}.

\begin{table}[t]
\begin{tabular}{|c|c|c|c|}
\hline 
Univ. class  & $\nu$                  &  $\eta$                & $\eta_q$   \\ \hline
O(2)         & 0.67169(7)             &  0.03810(8)            & 1.4722(2)  \\ \hline
O(4)         & 0.750(2)\phantom{00}   &  0.0360(3)\phantom{0}  & 1.371(1)\phantom{0}   \\ \hline
AH(25)       & 0.817(7)\phantom{00}   &    --                  & 0.882(2)\phantom{0}   \\ \hline
\end{tabular}
\caption{Critical exponents needed in the FSS analyses. For the
O(2) universality class we use $\nu$ and $\eta$ from Ref.~\cite{Hasenbusch:2019jkj} (where
$\omega=0.789(4)$ is also reported), see also Ref.~\cite{Chester:2019ifh}, and
$\eta_q=5-2Y_2$, with $Y_2$ from Ref.~\cite{HV:aniso} (to be used for the large $\kappa$ transition). 
For the O(4) universality class we use $\nu$ and $\eta$ from
Ref.~\cite{HV:aniso} (where $\omega\approx 0.79$ is also reported), see also
Ref.~\cite{Guida:1998bx}, and $\eta_q=5-2Y_2$, with $Y_2$ from
Ref.~\cite{HV:aniso} (to be used for the large $\kappa$ transition). 
For the Abelian Higgs universality class we use the results obtained in
Ref.~\cite{Bonati:2022oez} (see also Refs.~\cite{Bonati:2020jlm,
Bonati:2020ssr}); note that in this case the exponent $\eta$ is not defined
since the corresponding correlator is not gauge invariant and vanishes. 
For the O($N$) universality class in the large $N$ limit we use
$\nu=1-\frac{32}{3\pi^2 N}-\frac{32(27\pi^2-112)}{27\pi^4 N^2}$, 
see Refs.~\cite{Moshe:2003xn, Kos:2013tga}, and 
$\eta_q=1+\frac{64}{3\pi^2 N}-\frac{1024}{27\pi^4 N^2}$, 
see Refs.~\cite{Gracey:2002qa,Kos:2013tga}.}
\label{tab:critexp}
\end{table}

To identify the region $M_2M_3$ in Fig.~\ref{fig:phd_zq} we need to locate the
topological transitions departing from the $J=0$ and $J=\infty$ lines. Since
these transitions are not associated to any local order parameter, to detect
them we need to study the cumulants of the energy, and in particular the
third cumulant of the gauge part of the Hamiltonian $H_g$:
\begin{equation}
K_3=\langle H_g^3\rangle-3\langle H_g^2\rangle\langle H_g\rangle+2\langle H_g\rangle^3\ .
\end{equation} 
The use of the third (or higher) cumulant is particularly convenient to study
transitions with negative critical exponent $\alpha$, as the O(2)
ones\cite{Smiseth:2003bk}. Indeed the $n-$th cumulant satisfies the FSS relation
\begin{equation}
K_n=L^{n/\nu}[f_n(X)+O(L^{-\omega})]+L^3 K_{back}\ ,
\end{equation}
and $\alpha<0$ corresponds to $\frac{2}{\nu}<3$, thus the regular background
term $K_{back}$ dominates the FSS of the second cumulant in this case.

At first order phase transitions the specific heat and the Binder cumulant
develop peaks whose values scale linearly with the volume size~\cite{CLB-86,
VRSB-93}. For weak first order transitions this asymptotic behavior is however
often difficult to identify unambiguously, and it can be more convenient to
directly look for the emergence of a double peak structure in the energy
density. A different strategy, that is more effective in the case of a very
small latent heat, is to verify that the scaling relation Eq.~\eqref{eq:urxi},
typical of a second order phase transition, is
violated~\cite{Pelissetto:2019zvh}.

\subsection{The case $N=2$}

To investigate the ``small $N$'' case, we start by studying the phase diagram
of the $\mathbb{Z}_q^{(nc)}$ model with $q=2$, which is the first notrivial
value of $q$ (for $q=1$ scalars decouple).

\begin{figure}[t]
  \includegraphics*[width=0.95\columnwidth]{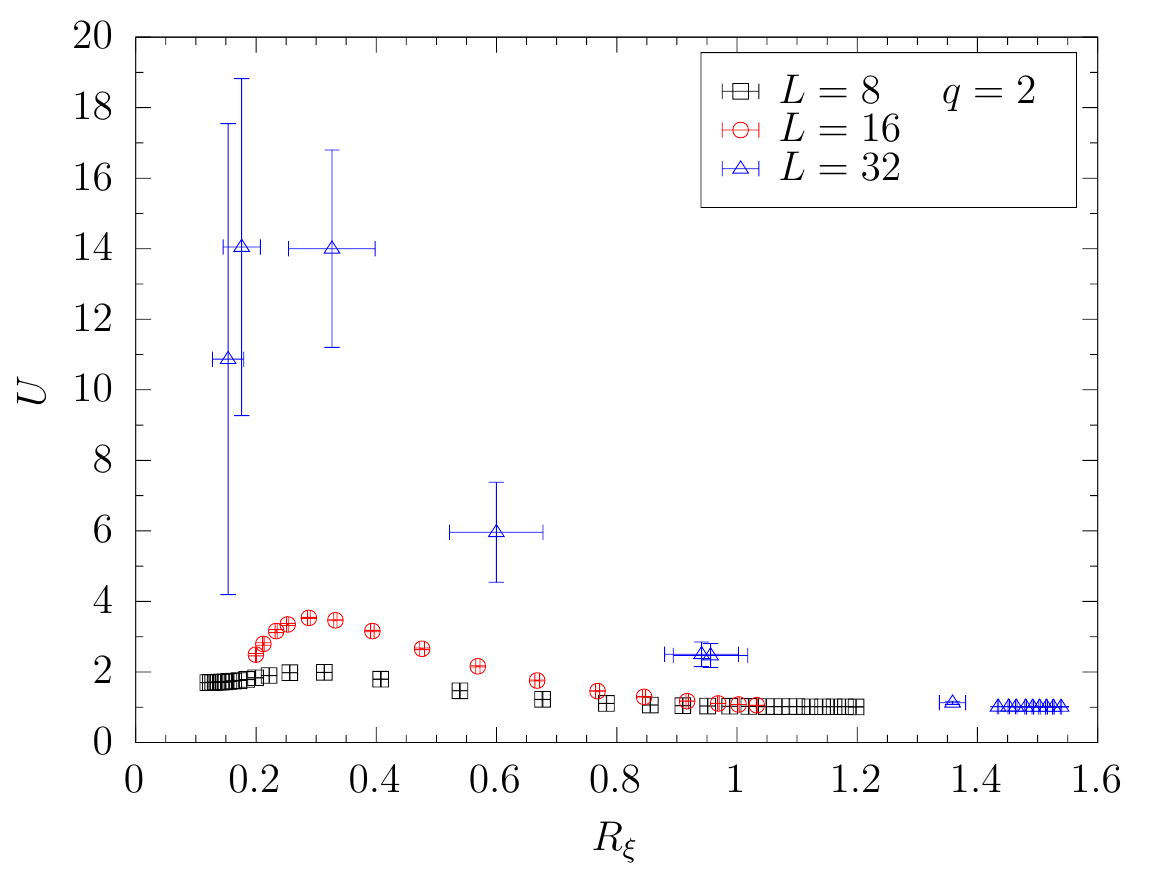}
  \caption{$N=2$, $q=2$, $\kappa=0.04$. Behaviour of $U$ as a function of
  $R_{\xi}$, obtained by varying the parameter $J$ in the Hamiltonian. 
  }
\label{fig:N2q2k0.04binder}
\end{figure}

\begin{figure}
  \includegraphics*[width=0.95\columnwidth]{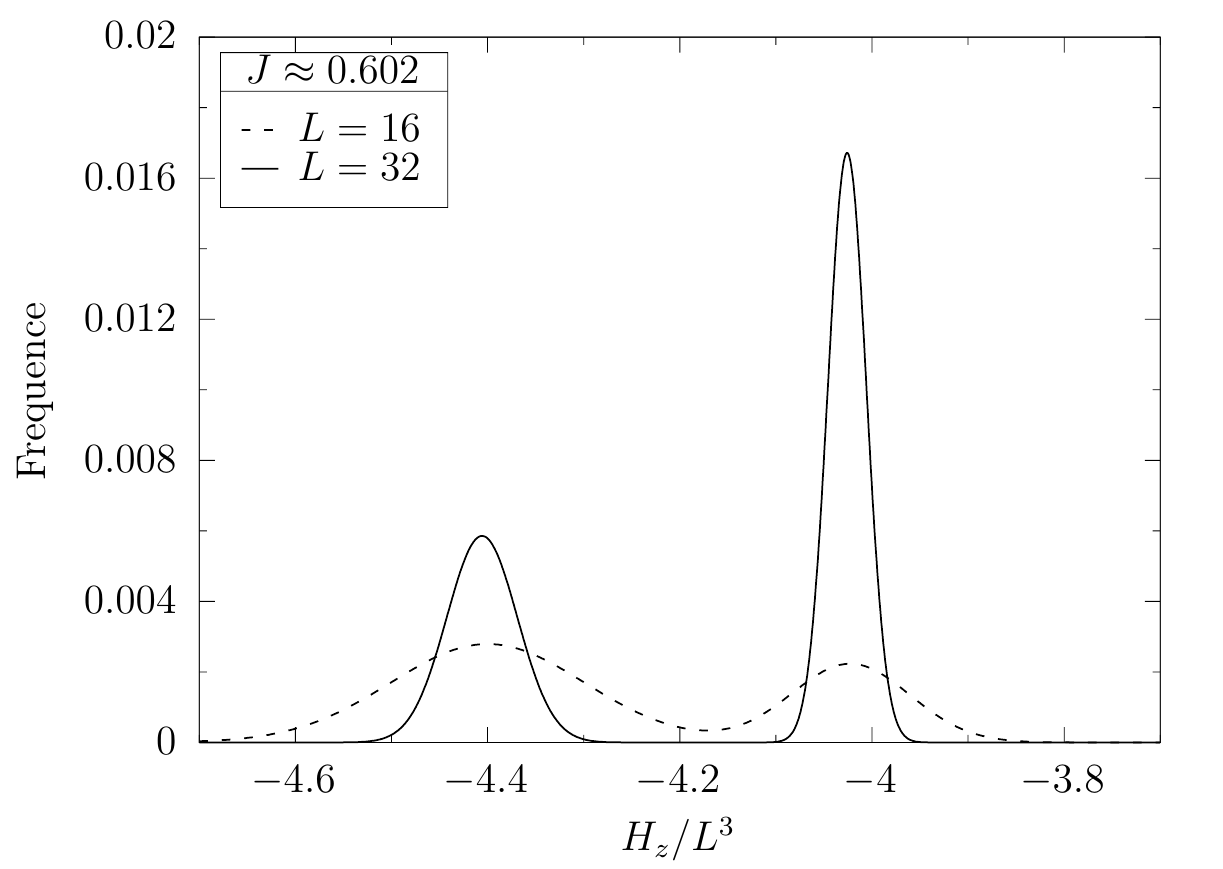}
  \caption{$N=2$, $q=2$, $\kappa=0.04$. Histograms of the 
  scalar part of energy density $H_z/L^3$ for $J\simeq 0.602$. 
  }
\label{fig:N2q2k0.04histo}
\end{figure}

To study the small $\kappa$ region we fix $\kappa=0.04$, a value smaller that
$\kappa_c^{\mathbb{Z}_q}(J=\infty)\approx 0.076$ in Eq.~\eqref{eq:ZqJinf}. By
varying $J$ we thus look for the presence of a phase transition using the
observables $R_{\xi}$ and $U$ introduced in Sec.~\ref{sec:obs}. A quite strong
first order transition is found for $J_c\approx 0.602$, 
with Monte Carlo metastabilities preventing a precise estimate of the critical coupling.
In Fig.~\ref{fig:N2q2k0.04binder} the behavior of $U$ as a function of
$R_{\xi}$ is reported, which show the diverging behavior typical of first
order phase transitions, with the sudden increase of the errorbars for $L=32$
being due to the appearance of long-lived metastable states. The first order
nature of this phase transition is also clear from the histograms of the scalar
part of energy density $H_z/L^3$, which are shown in
Fig.~\ref{fig:N2q2k0.04histo}. A double peak structure is present, which gets
more pronounced by increasing the lattice size.

\begin{figure}
  \includegraphics*[width=0.95\columnwidth]{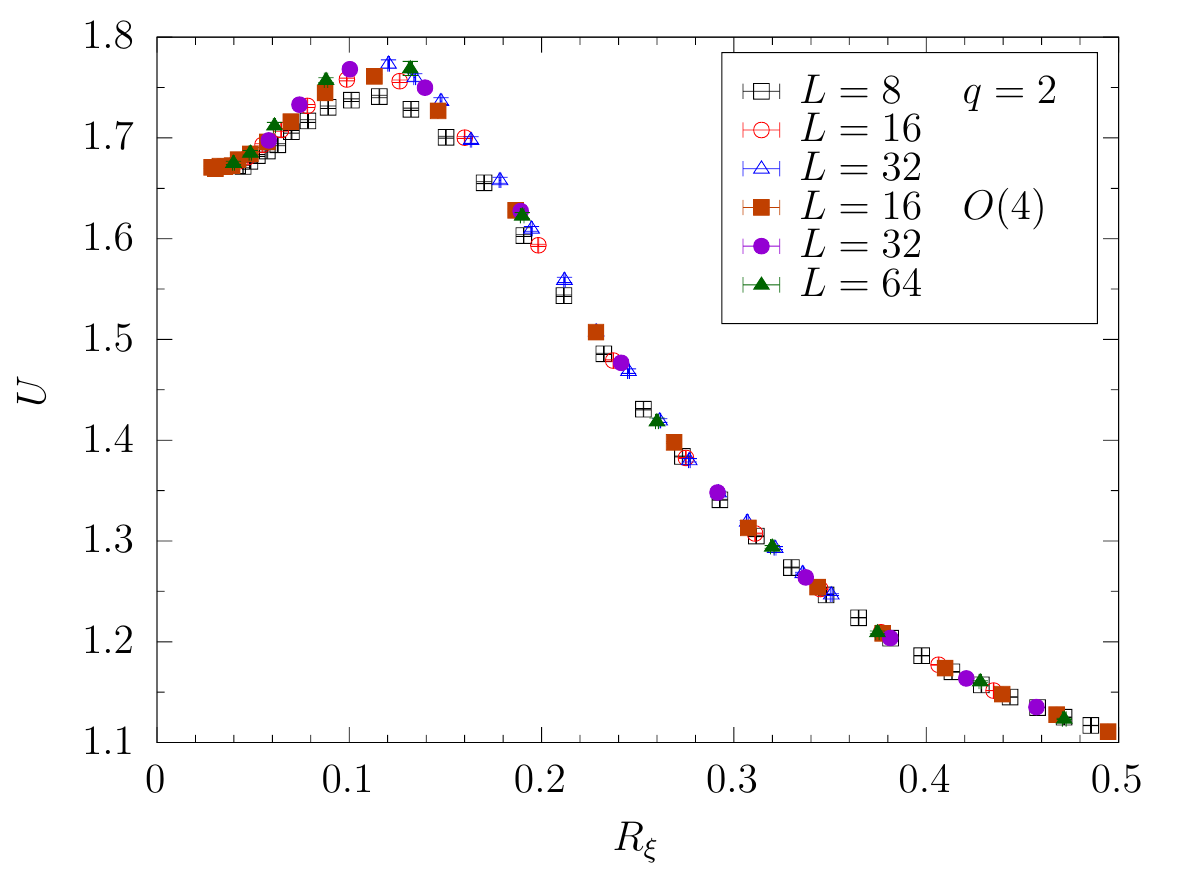}
  \caption{$N=2$, $q=2$, $\kappa=0.4$. Behaviour of $U$ as a function of
  $R_{\xi}$, obtained by varying the parameter $J$ in the Hamiltonian. 
  }
\label{fig:N2q2k0.4binder}
\end{figure}

We then move to the large $\kappa$ side of the phase diagram by fixing
$\kappa=0.4$, a value larger than $\kappa_c^{\mathbb{Z}_q}(J=0)\approx 0.3$
(see Eq.~\ref{eq:ZqJzero}). In this case a transition in the O(4) universality
class is found, as can be seen from Fig.~\ref{fig:N2q2k0.4binder}, where the
universal scaling curve obtained is compared to that of the O(4) model obtained
by fixing $A_{{\bm x},\mu}=0$.  Fitting the behavior of $R_{\xi}$ using the
known critical exponent $\nu$ of the O(4) model, see Tab.~\ref{tab:critexp}, we
obtain for the critical coupling the estimate $J_c=0.23433(5)$. This is only
slightly larger than the critical coupling $J_c^{\mathrm{O(4)}}=0.233965(2)$ of
the O(4) model, see Ref.~\cite{Ballesteros:1996bd} where the critical value of
$2NJ$ is reported.

\begin{figure}
  \includegraphics*[width=0.95\columnwidth]{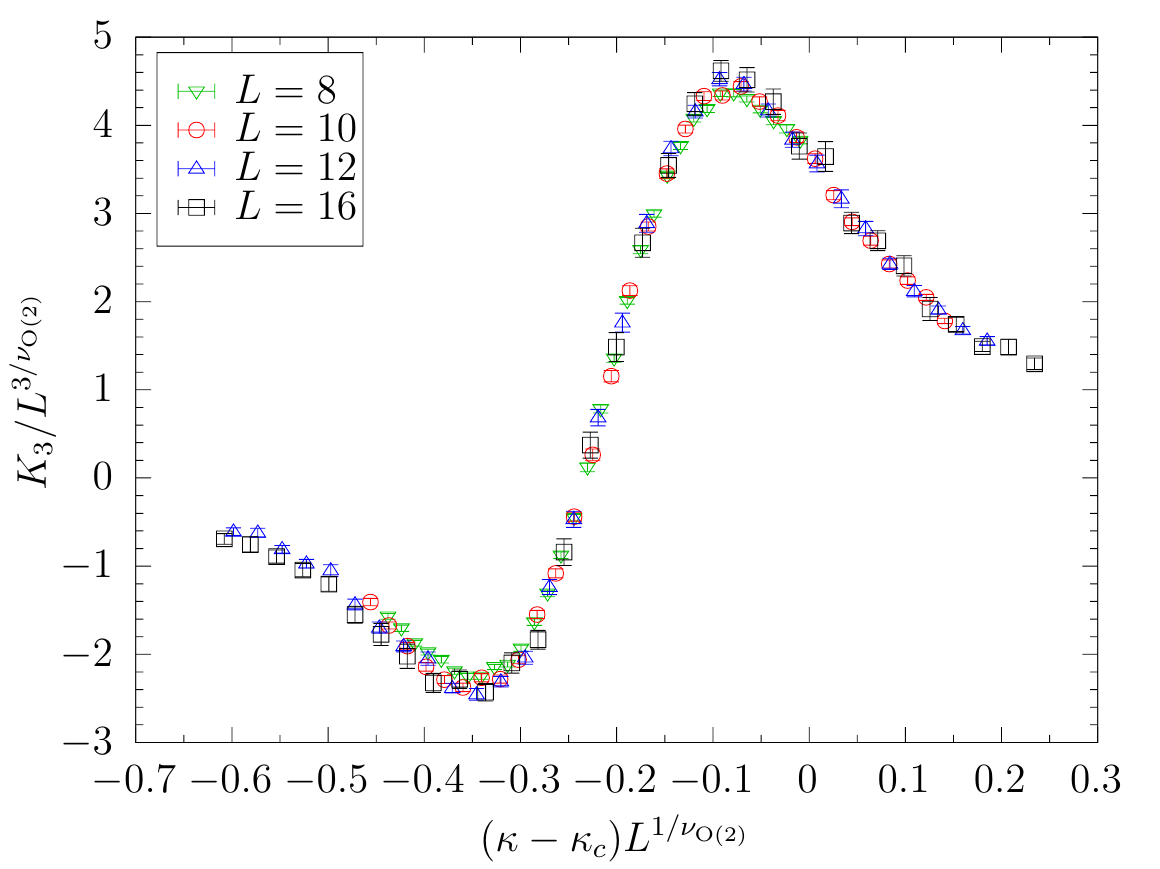} 
  \includegraphics*[width=0.95\columnwidth]{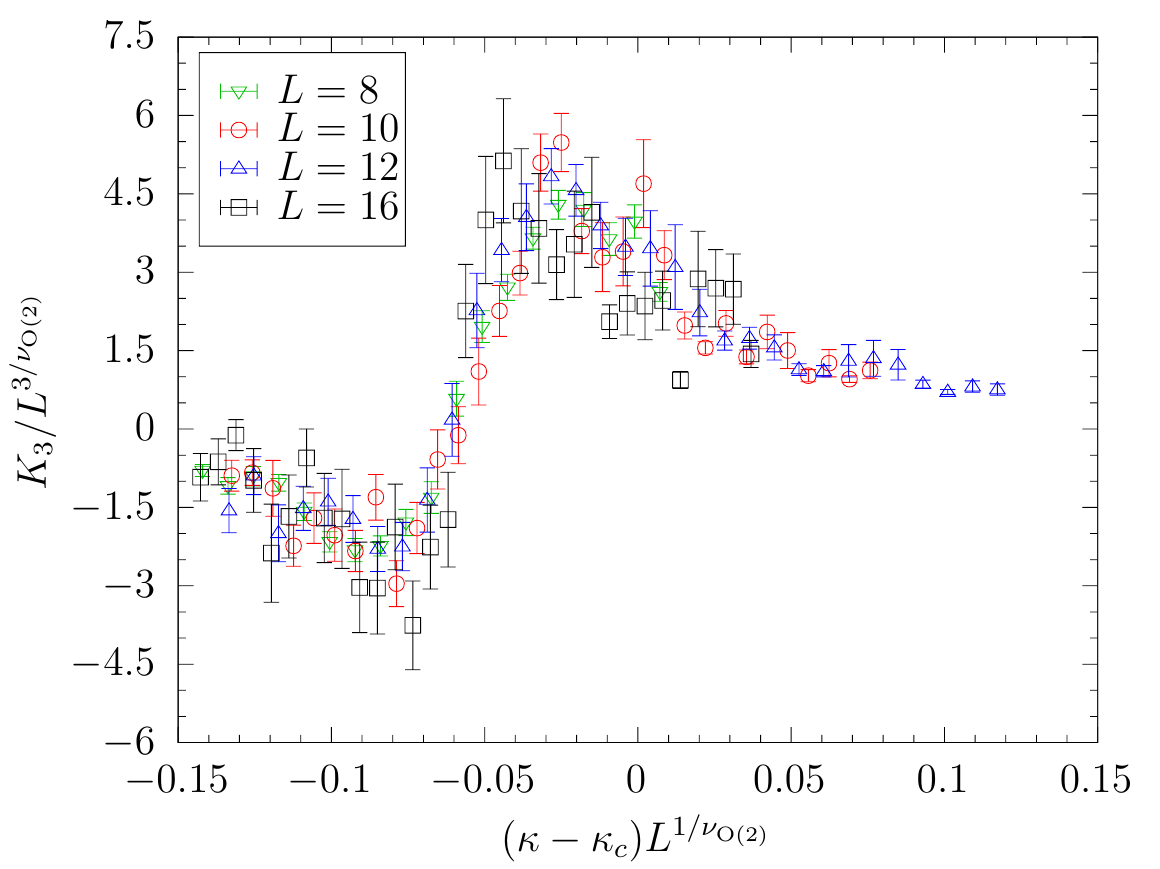} 
  \caption{$N=2$, $q=2$. Finite size scaling of the third cumulant $K_3$,
  obtained by using the known O(2) value of the critical exponent $\nu$
  (top) for $J=0.2$, with $\kappa_c=0.2998(7)$ (bottom) for $J=1$, with $\kappa_c=0.0763(4)$.
  }
\label{fig:N2q2J}
\end{figure}

To complete our preliminary scan of the phase diagram of the $q=2$ model, and
identify the $M_2M_3$ line in Fig.~\ref{fig:phd_zq}, we finally perform
simulations fixing $J=0.2$ (a value smaller than $J_c$ at $\kappa=0.4$ and
$\kappa=\infty$) and $J=1$ (a value larger than $J_c$ at $\kappa=0.04$). In
both the cases transitions of the O(2) universality class are found, as can be
seen from the FSS results shown in Fig.~\ref{fig:N2q2J}, obtained by using the
known value of the O(2) exponent $\nu$, see Tab.~\ref{tab:critexp}. For $J=0.2$
no scaling violations are observed and the transition is located at
$\kappa_c=0.2998(7)$; for $J=1$ scaling violations are sizable, and by
excluding the $L=8$ lattice data from the fit we estimate the critical coupling
to be $\kappa_c=0.0763(4)$. Both these values are quite
close to their asymptotic values for $J=0$ and $J=\infty$ respectively, see
Eqs.~\eqref{eq:ZqJinf}-\eqref{eq:ZqJzero}, signaling that the transition lines
emerging from the $J=0$ and $J=\infty$ sides of the phase diagram are almost
vertical.

\begin{figure}
  \includegraphics*[width=0.95\columnwidth]{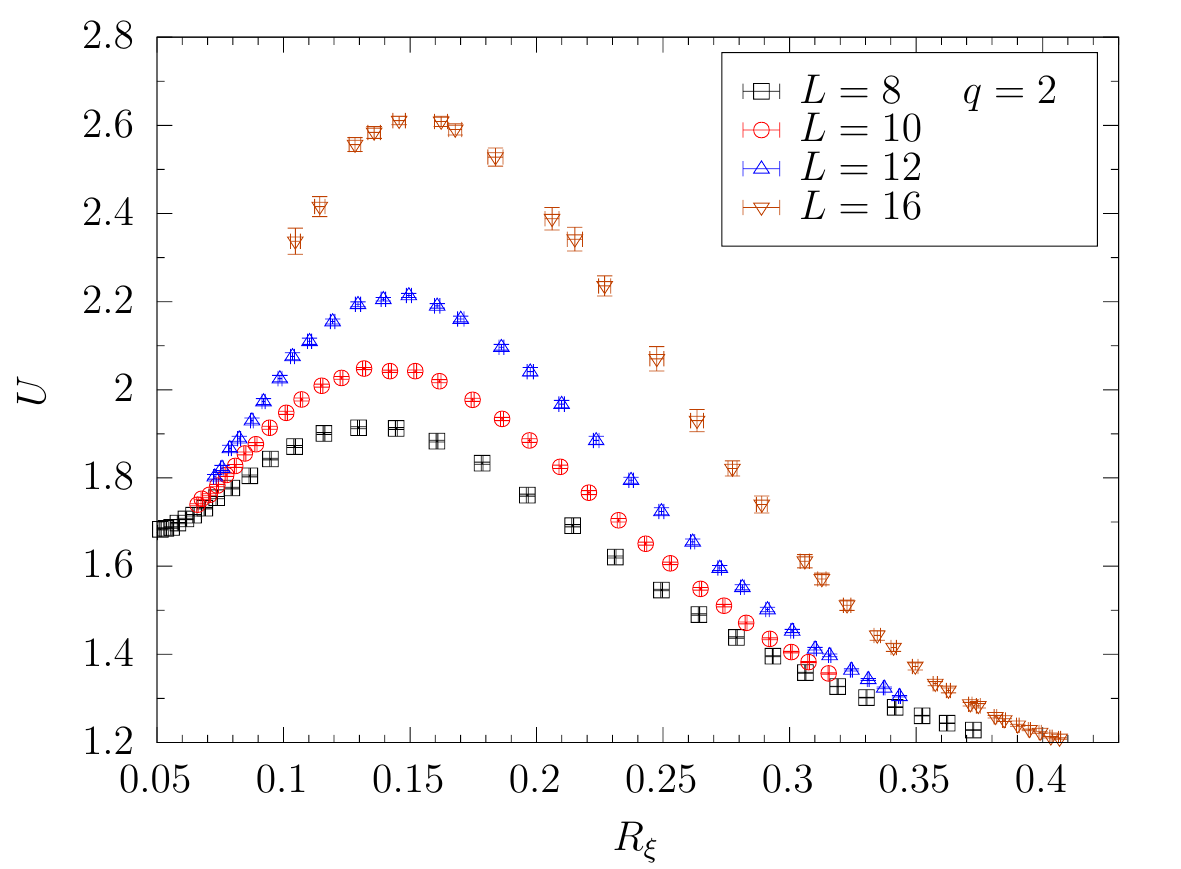}
  \caption{$N=2$, $q=2$, $\kappa=0.275$. Behaviour of $U$ as a function of
  $R_{\xi}$, obtained by varying the parameter $J$ in the Hamiltonian. 
  }
\label{fig:N2q2k0.275binder}
\end{figure}

\begin{figure}
  \includegraphics*[width=0.95\columnwidth]{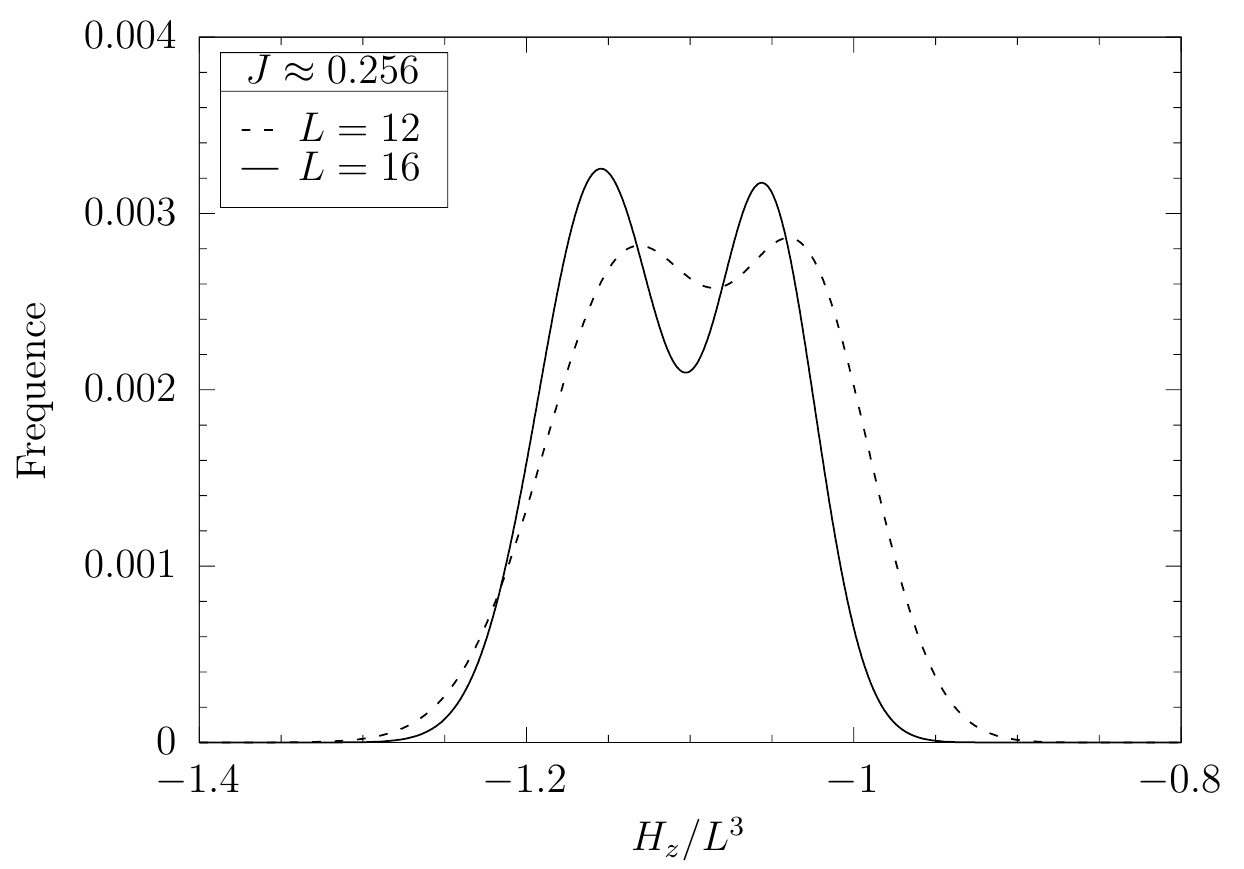}
  \caption{$N=2$, $q=2$, $\kappa=0.275$. Histograms of the 
  scalar part of energy density $H_z/L^3$ for $J\simeq 0.256$. 
  }
\label{fig:N2q2k0.275histo}
\end{figure}

We finally perform a simulation fixing $\kappa=0.275$, in order to cross the
$M_2M_3$ line in Fig.~\ref{fig:phd_zq}. The results obtained for $U$ as a
function of $R_{\xi}$ are shown in Fig.~\ref{fig:N2q2k0.275binder}: data
corresponding to different values of the lattice size $L$ do not collapse on
each other, and the peak values of $U$ at fixed $L$ increase significantly by
increasing $L$. We thus expect in this case the presence of a first order
transition, which is confirmed by the emergence of a double peak structure in
the energy density $H_z/L^3$ when increasing the lattice size, see
Fig.~\ref{fig:N2q2k0.275histo}. 

We can thus conclude that the phase diagram of the model with $q=2$ is fully
consistent with the one sketched in Fig.~\ref{fig:phd_zq}, with the possibly
interesting $M_2M_3$ line being a line of first order phase transitions. 

\begin{figure}
  \includegraphics*[width=0.95\columnwidth]{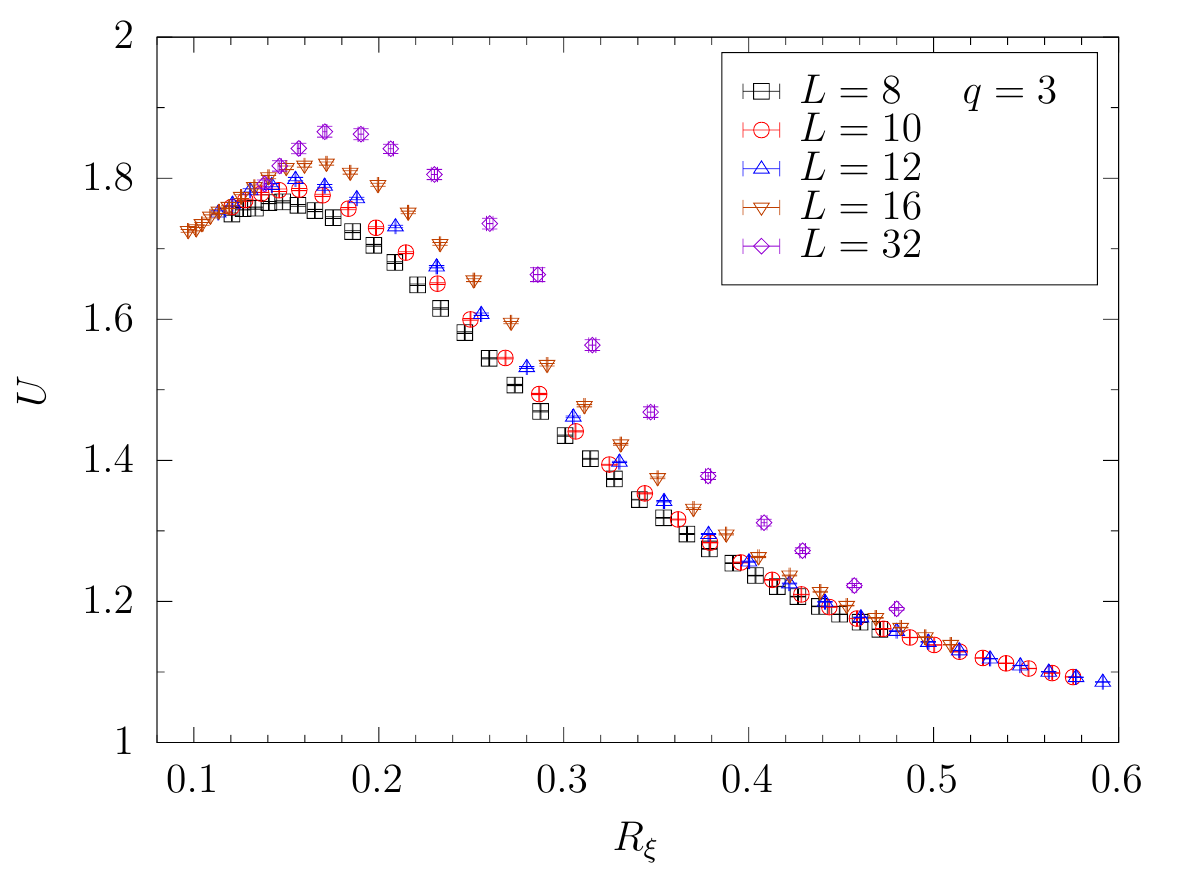}
  \includegraphics*[width=0.95\columnwidth]{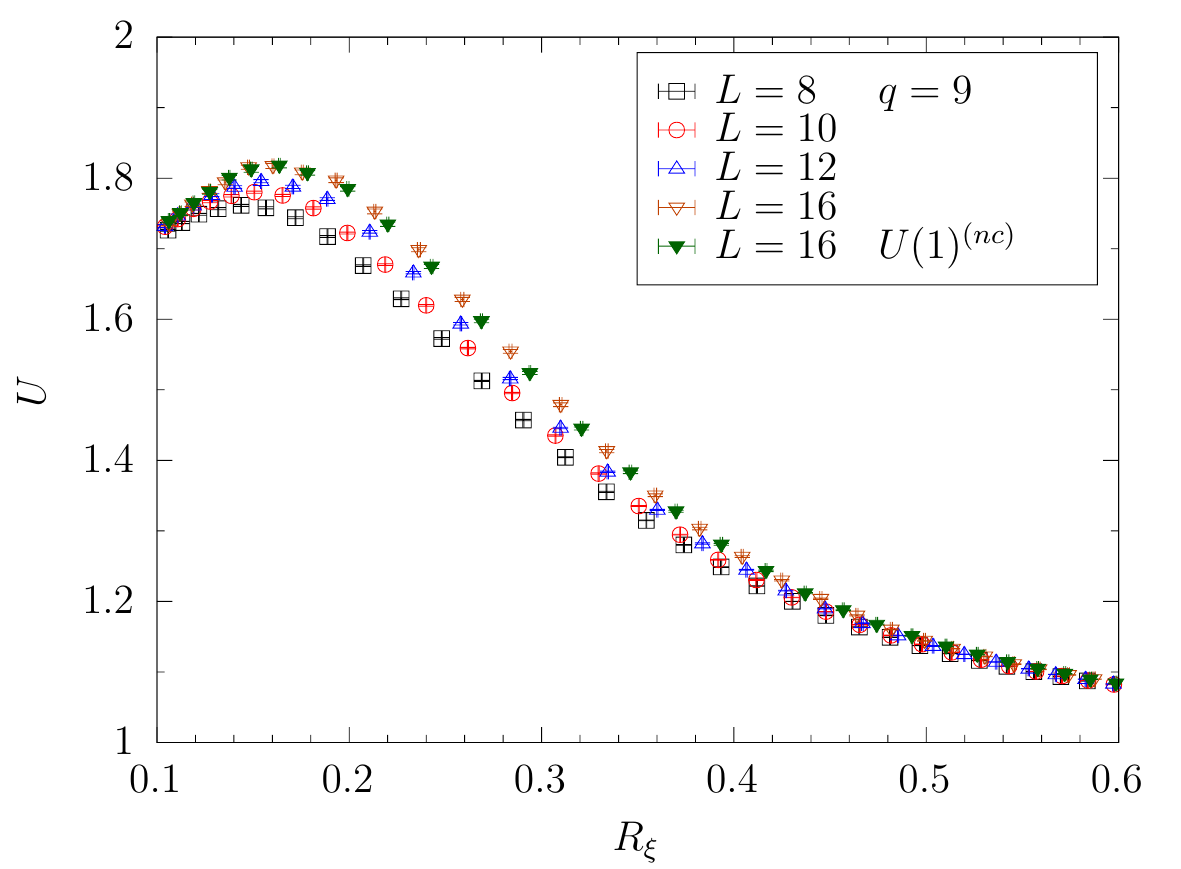}
  \caption{$N=2$, $\kappa=0.275$. Behaviour of $U$ as a function of $R_{\xi}$,
  obtained by varying the parameter $J$ in the Hamiltonian for the model with
  $q=3$ (top) and $q=9$ (bottom). In the latter case data for the U(1)$^{(nc)}$
  model are also shown for comparison. 
  }
\label{fig:N2q39}
\end{figure}

To close this section we present results obtained along the $M_2M_3$ line, always at
$\kappa=0.275$, for $q=3$ and $q=9$. In both the cases first order phase
transitions are found, as seen from Figs.~\ref{fig:N2q39}: no scaling is
observed in the $U$ vs $R_{\xi}$ plot, however the strength of the first order
transition decreases when increasing $q$, and for $q=9$ data are practically
indistinguishable from those of the U(1)$^{(nc)}$ model, in which a very weak
first order phase transition is present for $N=2$, see Refs.~\cite{KMPST-08,
Bonati:2020jlm}.

\subsection{The case $N=25$}

We now discuss the results obtained for the model with $25$ scalar flavours,
starting again from the $q=2$ gauge discretization and focussing on the most
interesting part of the phase diagram.

\begin{figure}
  \includegraphics*[width=0.95\columnwidth]{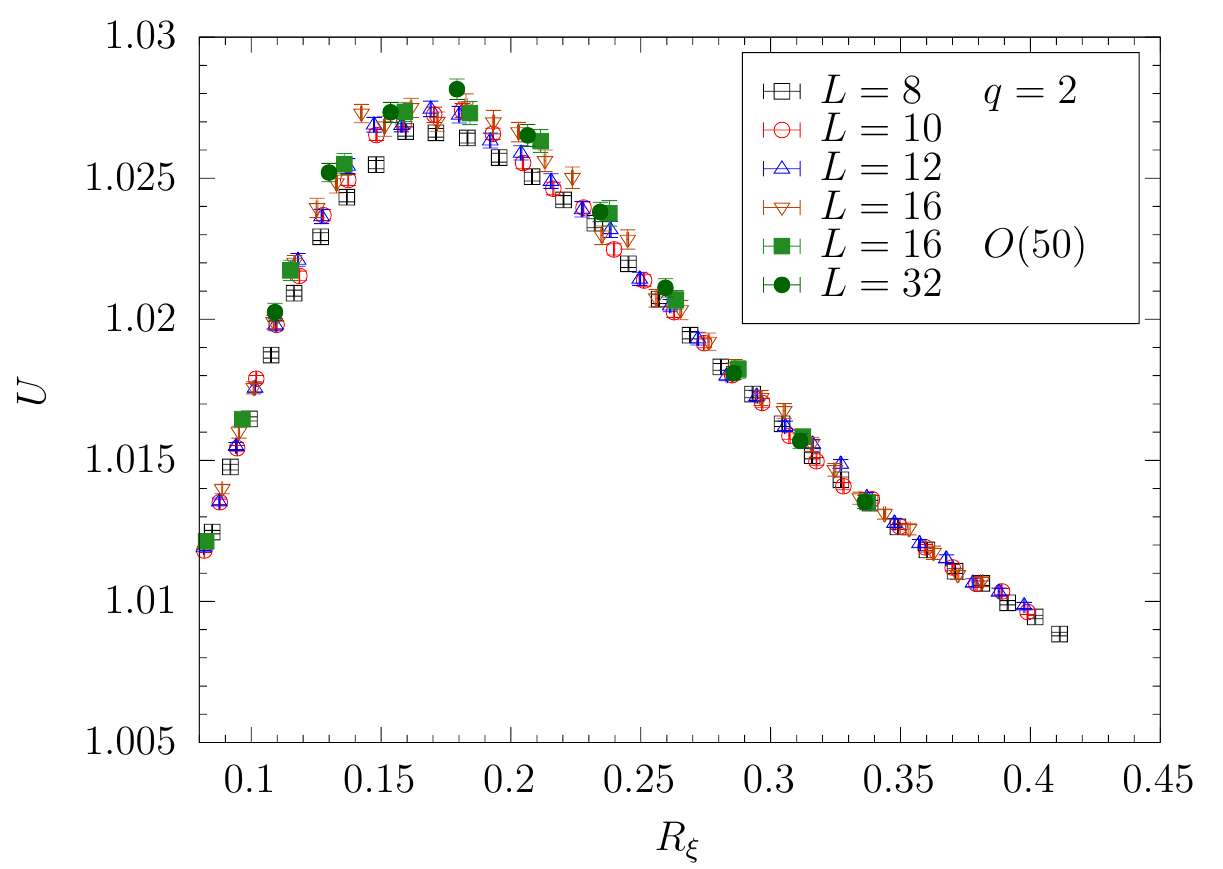}
  \caption{$N=25$, $q=2$, $\kappa=0.4$. Behaviour of $U$ as a function of
  $R_{\xi}$, obtained by varying the parameter $J$ in the Hamiltonian. 
  }
\label{fig:N25q2k0.4binder}
\end{figure}

\begin{figure}
  \includegraphics*[width=0.95\columnwidth]{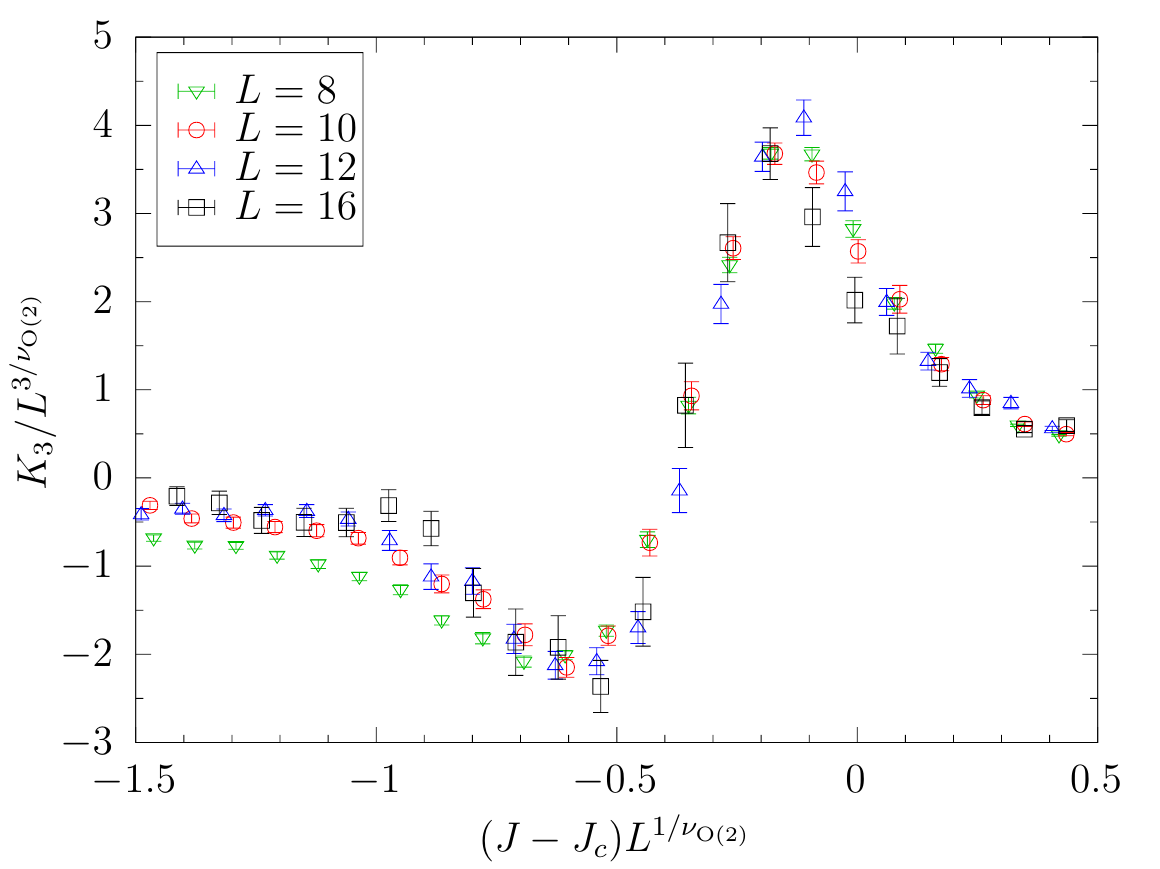}
  \includegraphics*[width=0.95\columnwidth]{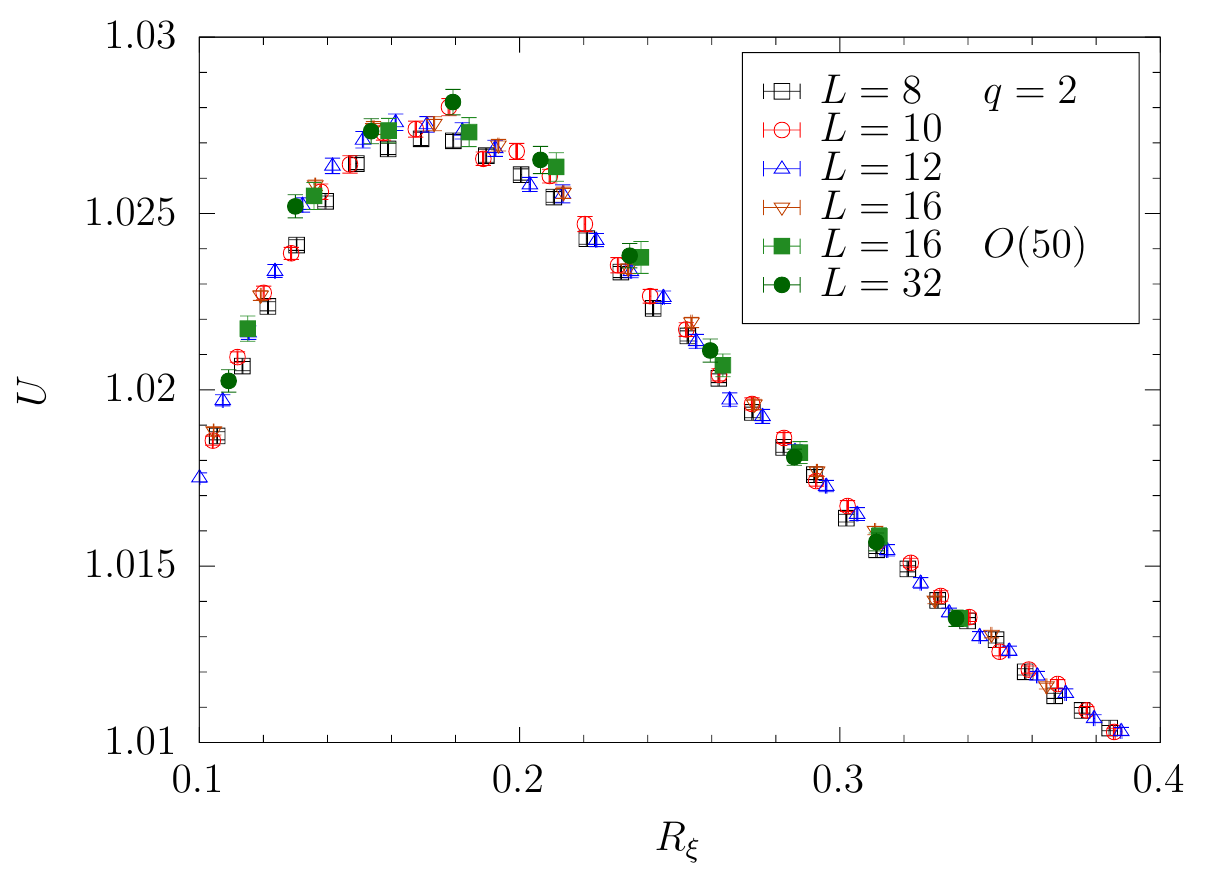}
  \caption{$N=25$, $q=2$, $\kappa=0.275$. 
  (top) Finite size scaling of the third cumulant $K_3$, obtained by using the
  known O(2) value of the critical exponent $\nu$ and $J_c=0.181(2)$.  
  (bottom) Behaviour of $U$ as a function of $R_{\xi}$, obtained by varying the
  parameter $J$ in the Hamiltonian. 
  }
\label{fig:N25q2k0.275}
\end{figure}

We first of all present the results obtained for $\kappa=0.4$ (larger than
$\kappa^{\mathbb{Z_q}}_c(J=0)\approx 0.3$, see Eq.~\eqref{eq:ZqJzero}), where a
transition of the O(50) universality class is expected. Results reported in
Fig.~\ref{fig:N25q2k0.4binder} are fully consistent with this expectation,
since the scaling curve obtained for $U$ against $R_{\xi}$ is well compatible
with the one of the O(50) model, determined by fixing $A_{{\bm x},\mu}\equiv 0$
in the simulations. To fit the behavior of $R_{\xi}$ we use the large $N$
prediction of $\nu$ reported in the caption of Tab.~\ref{tab:critexp},
obtaininig the estimate $J_c=0.2502(3)$ for the critical coupling.  This value
is already quite close to the asymptotic large $N$ critical coupling of the
O(N) models, $J_c=0.252731\ldots$, reported in Ref.~\cite{Campostrini:1995np}.

\begin{figure}
  \includegraphics*[width=0.95\columnwidth]{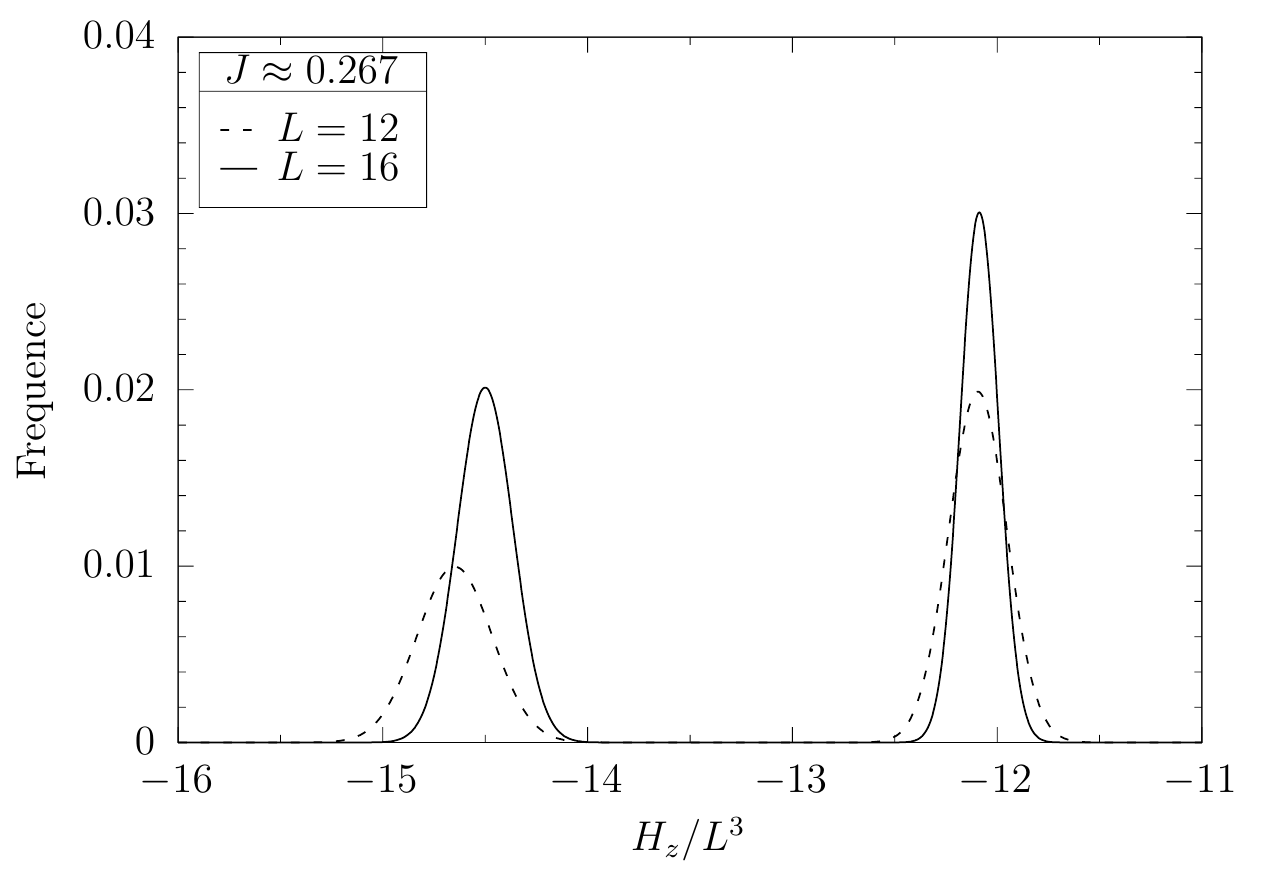}
  \caption{$N=25$, $q=3$, $\kappa=0.275$. 
  Histograms of the 
  scalar part of energy density $H_z/L^3$ for $J\simeq 0.267$. 
  }
\label{fig:N25q3k0.275histo}
\end{figure}

\begin{figure}
  \includegraphics*[width=0.95\columnwidth]{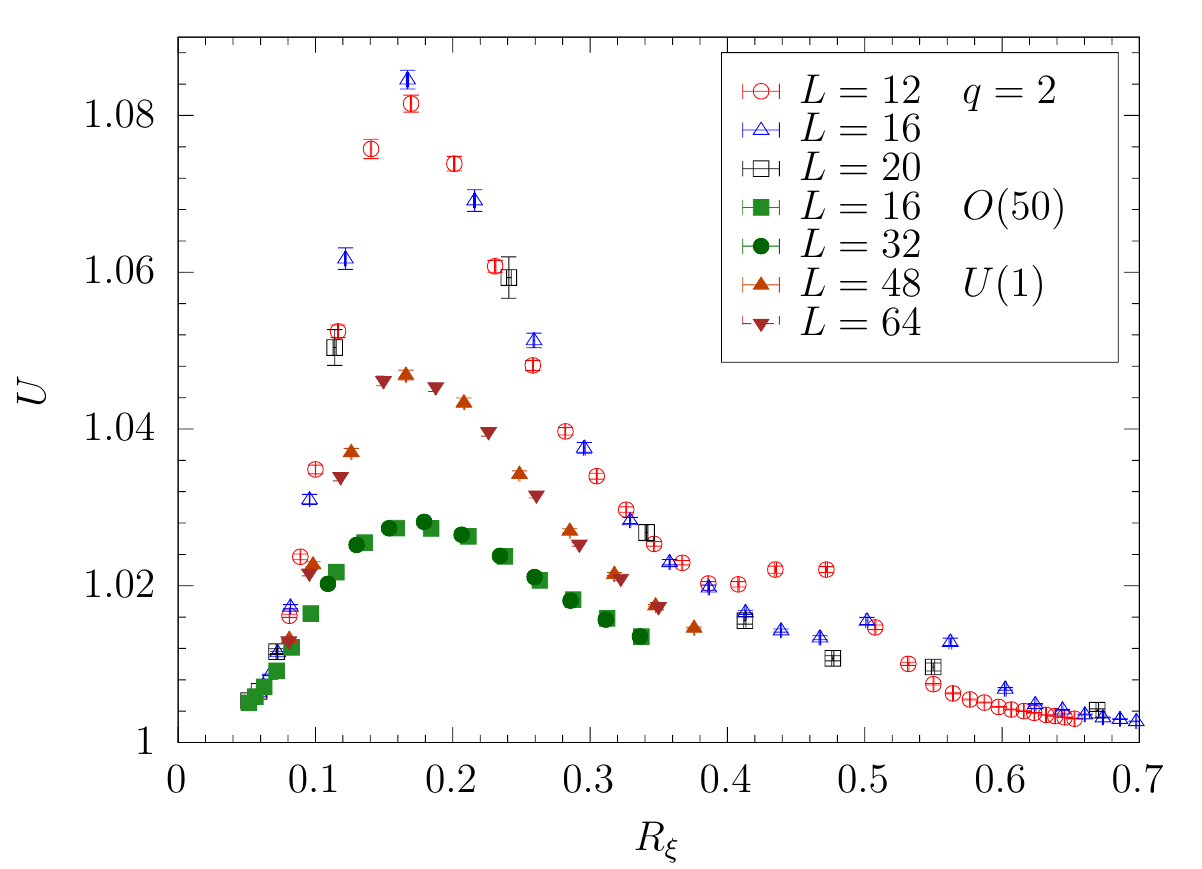}
  \caption{$N=25$, $q=4$, $\kappa=0.275$. Behaviour of $U$ as a function of
  $R_{\xi}$, obtained by varying the parameter $J$ in the Hamiltonian. For
  comparison data for the O(50) and the U(1)$^{(nc)}$ models 
  from Ref.~\cite{Bonati:2020jlm} are also reported.
  }
\label{fig:N25q4k0.275binder}
\end{figure}

Other simulations have been performed at $\kappa=0.275$, which is smaller than
$\kappa^{\mathbb{Z}_q}_c(J=0)\approx 0.3$. However for $N=25, q=2$ the
transition line emerging from the $J=0$ critical point is not vertical anymore,
and in this case we found two transitions: an O(2) transition at $J_c=0.181(2)$
and an O(50) transition at $J_c=0.2506(3)$, as can be seen from the FSS curves
shown  in Fig.~\ref{fig:N25q2k0.275}. To approximately locate the position of
the multicritical point $M_3$ in Fig.~\ref{fig:phd_zq}, we thus performed
simulations at fixed $J=0.2$, finding an O(2) transition at
$\kappa_c=0.2554(15)$.  Simulations at fixed $\kappa=0.2$ show evidence of two
very close transitions at $J_c\approx 0.2518$, providing our best estimate for
the position of the multicritical point $M_3$. Finally, simulations performed
at $\kappa=0.15$ found a first order transition at $J\approx 0.25$, with hints
of a continuous transition for slightly larger values of the coupling $J$; it
is thus possible that this point is on the left of the multicrical point $M_2$
in Fig.~\ref{fig:phd_zq}, or anyway very close to it. 

The region $M_2M_3$ is thus quite small for $N=25$, $q=2$,
and significant crossover effects are expected to be found due to the nearby
O(50) and first order transition lines. Since a complete investigation of the
small $q$ case is not our principal aim, we leave a detailed analysis of
this region of the parameter space to future studies.

The model with $N=25$, $q=3$ is much simpler: in this case
$\kappa_c^{\mathbb{Z}_q}(J=0)\approx 0.68$ (see Eq.~\ref{eq:ZqJzero}), and by
performing simulations at $\kappa=0.7$ a clear O(50) transition is found for
$J_c= 0.25051(15)$. However simulations at $\kappa=0.275$ provide clear
evidence of a strong first order phase transition for $J_c\simeq 0.267$, see
the histograms reported in Fig.~\ref{fig:N25q3k0.275histo}. 

The interpretation of the case $N=25$, $q=4$ is again problematic: now
$\kappa_c^{\mathbb{Z}_q}(J=0)\approx 1.2$ (see Eq.~\ref{eq:ZqJzero}), but the
results of simulations performed both at $\kappa=0.4$ and $\kappa=0.275$ do not
provide clear indications on the nature of the critical behavior. In both the
cases very large correction to scaling are found, with data for $U$ against
$R_{\xi}$ that seem to approach an asymptotic curve that does not correspond
neither to the Abelian Higgs nor the O(50) universality classes, see
Fig.~\ref{fig:N25q4k0.275binder} for the $\kappa=0.275$ case. To make things
worst, the apparent asymptotic curve of the $\kappa=0.275$ data is different
from the one obtained for $\kappa=0.4$. The most natural interpretation of
these results is that much larger lattices would be needed to really resolve the
true critical behavior of the model. 

\begin{figure}
  \includegraphics*[width=0.95\columnwidth]{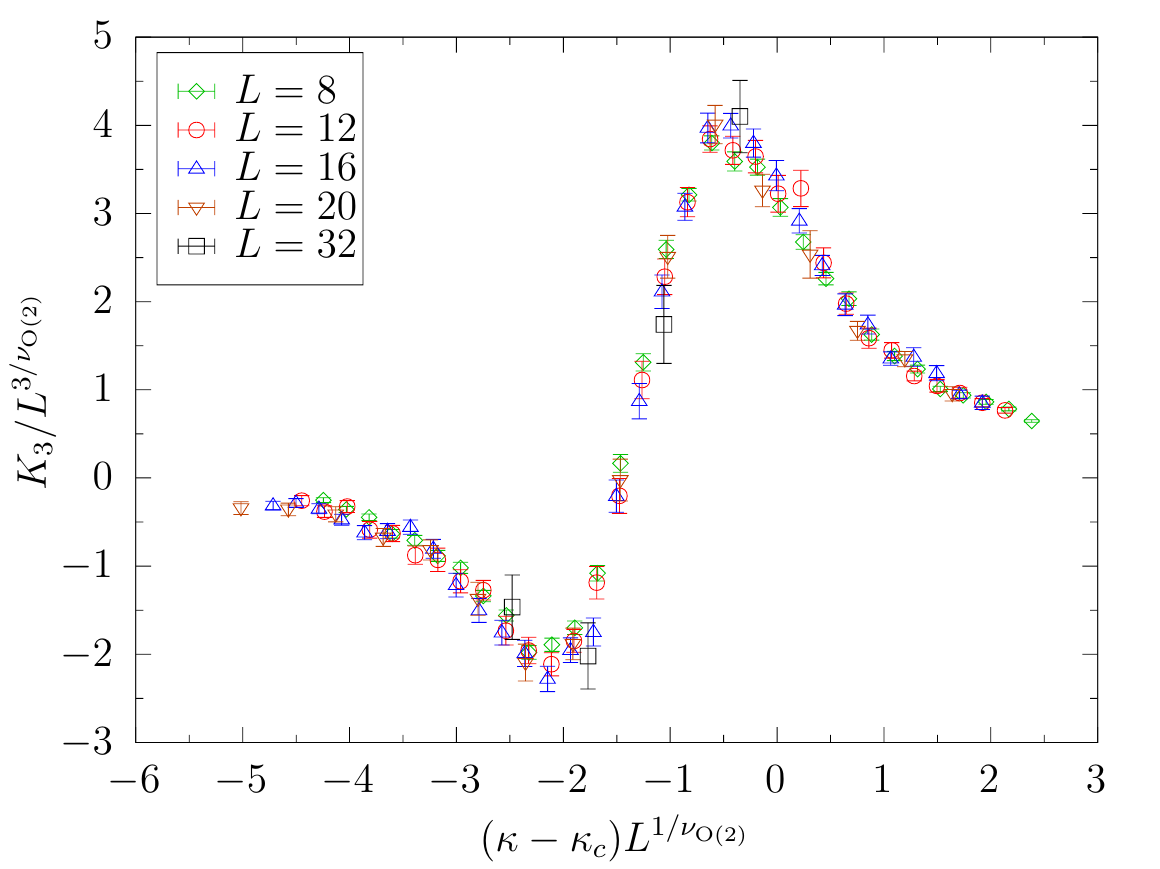}
  \caption{$N=25$, $q=5$, $J=0.2$. 
  Finite size scaling of the third cumulant $K_3$, obtained by using the
  known O(2) value of the critical exponent $\nu$ and $\kappa_c=1.792(1)$.
  }
\label{fig:N25q5J0.2}
\end{figure}

\begin{figure}
  \includegraphics*[width=0.95\columnwidth]{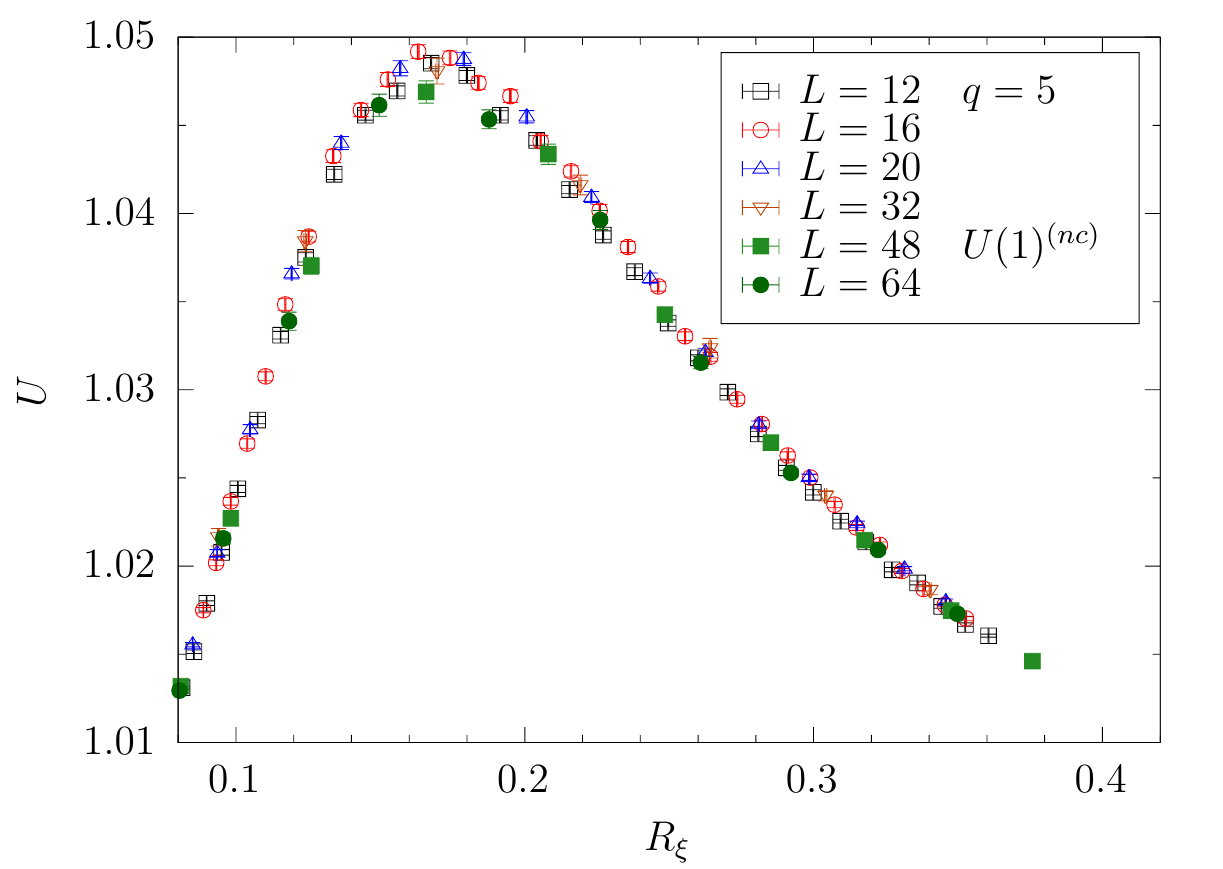}
  \caption{$N=25$, $q=5$, $\kappa=0.4$. Behaviour of $U$ as a function of
  $R_{\xi}$, obtained by varying the parameter $J$ in the Hamiltonian. For
  comparison data for the U(1)$^{(nc)}$ model from Ref.~\cite{Bonati:2020jlm} are also
  reported.
  }
\label{fig:N25q5k0.4}
\end{figure}

\begin{figure}
  \includegraphics*[width=0.95\columnwidth]{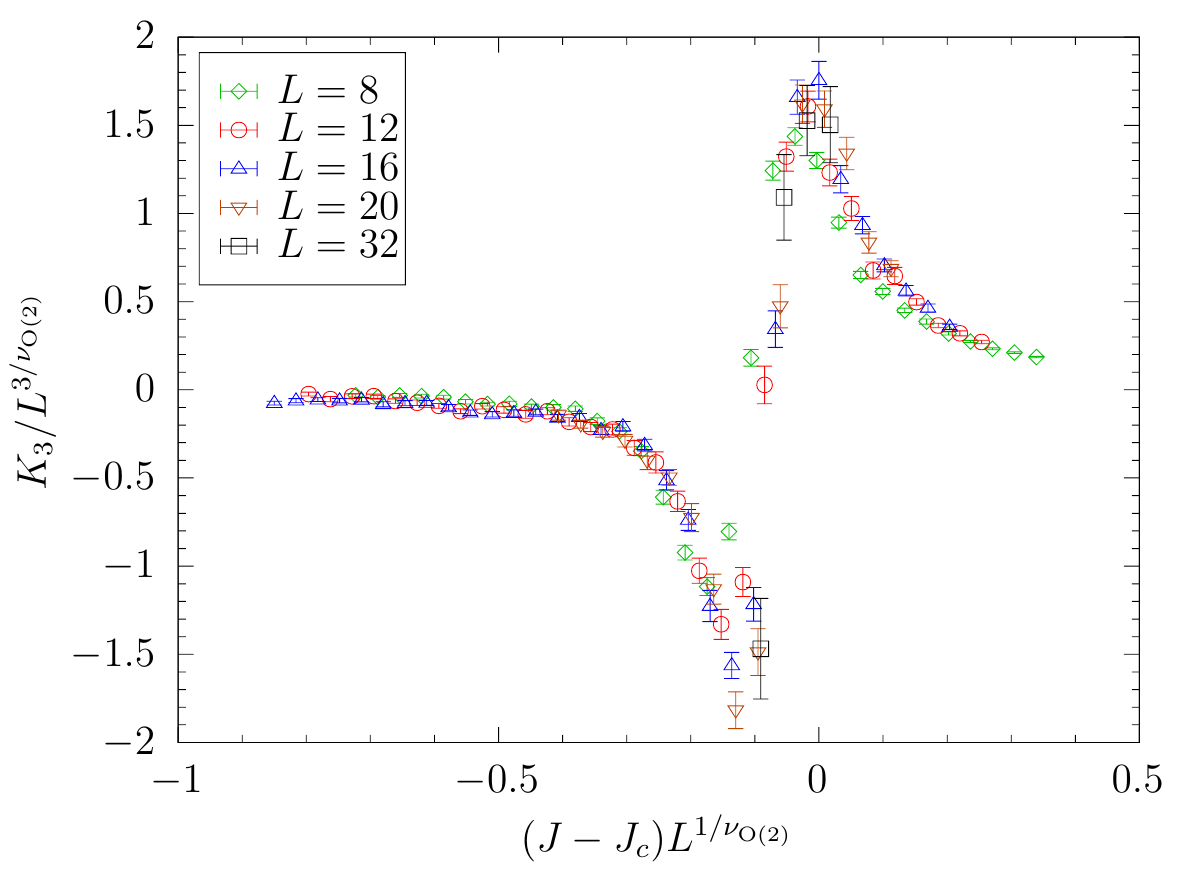}
  \caption{$N=25$, $q=5$, $\kappa=1.2$. 
  Finite size scaling of the third cumulant $K_3$, obtained by using the
  known O(2) value of the critical exponent $\nu$ and $J_c=0.2727$
  ($L=8$ was not included in the fit to determine $J_c$).
  }
\label{fig:N25q5k1.2cum}
\end{figure}

\begin{figure}
  \includegraphics*[width=0.95\columnwidth]{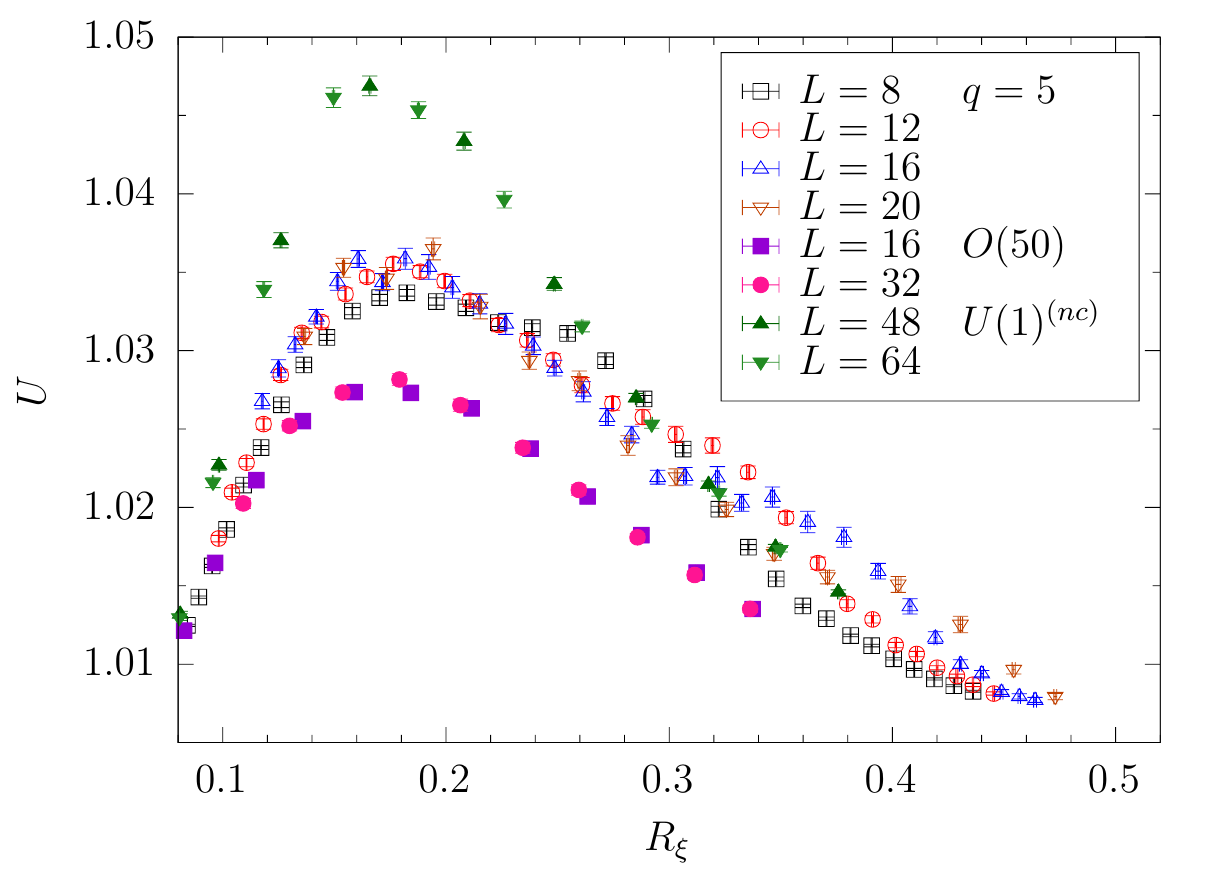}
  \caption{$N=25$, $q=5$, $\kappa=1.2$. Behaviour of $U$ as a function of
  $R_{\xi}$, obtained by varying the parameter $J$ in the Hamiltonian. For
  comparison data for the O(50) and U(1)$^{(nc)}$ models from Ref.~\cite{Bonati:2020jlm} 
  are also reported.
  }
\label{fig:N25q5k1.2binder}
\end{figure}

\begin{figure}
  \includegraphics*[width=0.95\columnwidth]{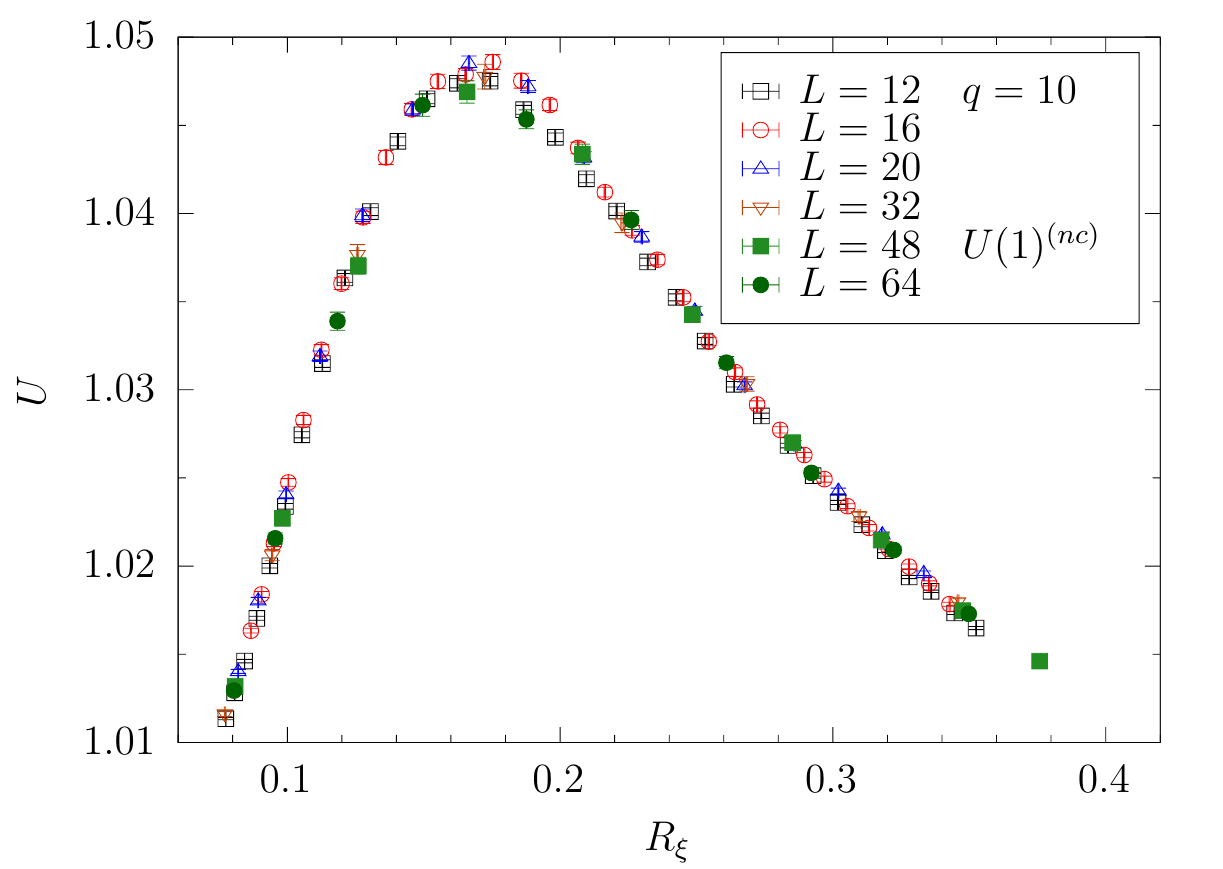}
  \caption{$N=25$, $q=10$, $\kappa=0.4$. Behaviour of $U$ as a function of
  $R_{\xi}$, obtained by varying the parameter $J$ in the Hamiltonian. For
  comparison data for the U(1)$^{(nc)}$ model from Ref.~\cite{Bonati:2020jlm} 
  are also reported.
  }
\label{fig:N25q10k04binder}
\end{figure}

The model with $N=25$, $q=5$ turns out to be the most interesting one.  For
this model $\kappa_c^{\mathbb{Z}_q}(J=0)\approx 1.9$ (see
Eq.~\ref{eq:ZqJzero}), and by performing simulations at $J=0.2$ and $J=0.25$ we
find clear O(2) transitions at $\kappa_c=1.792(1)$ and $\kappa_c=1.509(2)$
respectively, see Fig.~\ref{fig:N25q5J0.2} for the case $J=0.2$. Fixing
$\kappa=0.4$ and scanning in the coupling $J$ we find the symmetry enlargement
we were looking for: the universal scaling curve of $U$ against $R_{\xi}$ is
indeed the same as that of the U(1)$^{(nc)}$ model, as can be appreciated from
data reported in Fig.~\ref{fig:N25q5k0.4}. By using the critical exponent
$\nu$ reported in Tab.~\ref{tab:critexp} for the Abelian Higgs universality
class, we obtain for the critical coupling the estimate $J_c=0.29509(2)$, which
is already remarkably close to the critical coupling $J_c^{U(1)}=0.295511(4)$ of the
U(1)$^{(nc)}$ model with $N=25$ for $\kappa=0.4$ (see Ref.~\cite{Bonati:2020jlm}).

Simulations of the $N=25$, $q=5$ model have been carried out also for
$\kappa=1.2$, which turned out to be quite close to the multicritical point
$M_3$ in Fig.~\ref{fig:phd_zq}. Two nearby transitions can indeed be found at
$J_c\approx 0.2674$ and $J_c\approx 0.2727$, detected by using $R_{\xi}$ and
$U$, and $K_3$ respectively. The scaling of $K_3$ at the transition with
$J_c\approx 0.2727$ is consistent with the exponents of the O(2) universality
class, see Fig.~\ref{fig:N25q5k1.2cum} ($L=8$ was not included in the fit for
find $J_c$). The scaling of $U$ against $R_{\xi}$ at $J_c\approx 0.2674$ is
instead nontrivial, as can be seen from Fig.~\ref{fig:N25q5k1.2binder}.  Data
seems to collapse on a common scaling curve, although significant corrections
to scaling are present, especially in the right part of the figure, where a
contamination coming from the second transition is present. The significant
thing to note is that this scaling curve is however different from universal
curves of the O(50) and of the U(1)$^{(nc)}$ models, also shown in
Fig.~\ref{fig:N25q5k1.2binder}. This behavior can be explained in a natural way
by assuming the multicritical point $M_3$ to be associated to a continuous
transition, whose scaling function is the one 
on which data points in Fig.~\ref{fig:N25q5k1.2binder} collapse, due to a crossover phenomenon.

Finally, to verify that the symmetry enlargement observed for $q=5$ is present
also for larger values of the discretization parameter, we present results
obtained for the model with $q=10$, again for $\kappa=0.4$.  As expected, also
in this case the symmetry enlargement to the Abelian Higgs universality class
is present, as can be seen from Fig.~\ref{fig:N25q10k04binder}. In this case
the transition is located at $J_c=0.29555(2)$ which is only two standard
deviations away from the value $J_c^{U(1)}=0.295511(4)$ obtained in
Ref.~\cite{Bonati:2020jlm} in the U(1)$^{(nc)}$ model.

\section{Conclusions}
\label{sec:concl}

In this work we studied a variant of the non-compact multicomponent lattice
Abelian Higgs model with reduced gauge symmetry, with the aim of investigating
whether the discrete $\mathbb{Z}_q^{(nc)}=2\pi\mathbb{Z}/q$ gauge symmetry is sufficient for the
model to display transitions in the continuous Abelian Higgs universality
class.

In studying this model we considered two different values for the number of
scalar flavors, namely $N=2$ and $N=25$. Although the topology of the phase
diagram is the same in these two cases, the universality classes of the
transitions present in these two cases are very different.
Indeed the results obtained in the model with gauge symmetry
$\mathbb{Z}_q^{(nc)}$ are expected to converge, for large $q$, to those of the
model with gauge symmetry U(1)$^{(nc)}$, and only for large enough $N$ the
U(1)$^{(nc)}$ model exhibit transitions in which both gauge and scalar degrees
of freedom become critical~\cite{MV-08, KMPST-08, Bonati:2020jlm}.

We thus verified that for $N=2$ the numerical results are consistent with the
absence of any symmetry enlargement, since both the $\mathbb{Z}_q^{(nc)}$ and
the U(1)$^{(nc)}$ gauge theories display first order phase transitions in large
parts of the phase diagram.

The case $N=25$ is clearly the most interesting one. The analysis of the values
$q=2$ and $q=4$ of the gauge discretization parameter can not be considered
conclusive, since large crossover effects seem to be present. For
$q\ge 5$, instead, we unambiguously identified regions of the parameter space in which
the $\mathbb{Z}_q^{(nc)}$  gauge symmetry enlarges to U(1)$^{(nc)}$, and
the model with discrete gauge group exhibits transitions of the continuous
Abelian Higgs universality class.

This is not incompatible with the negative results recently obtained in
Ref.~\cite{discretecompact}, where an analogous discretization of the compact
Abelian Higgs model with charge $Q=2$ has been studied, since the presence of
first order phase transitions can never be excluded by universality arguments
alone.  However, it will be surely interesting to understand, in future
studies, the dynamical origin of this difference, to better understand the
relation between the compact and the non-compact models~\cite{Bonati:2020jlm,
Bonati:2020ssr, Bonati:2022oez}. In particular, it is still an open question whether
transitions of the continuous Abelian Higgs universality class are possible in
a lattice model with a finite Abelian gauge group, like the one studied in
Ref.~\cite{discretecompact} but unlike the one used in the present work (which
is discrete but infinite).

We finally note that the results obtain at $N=25$, $q=5$ for $\kappa=1.2$
suggest the multicritical point $M_3$ in Fig.~\ref{fig:phd_zq} to be associated
to a continuous phase transition. This is something that surely deserves to be
further investigated, both from the numerical and from the analytical point of
view. Such a continuous transition would indeed correspond to a very peculiar
multicritical theory, with lines of O(2N), Abelian Higgs and O(2) (ordinary
and topological) transitions crossing each other.

\emph{Acknowledgement}. Numerical simulations have been performed on the CSN4
cluster of the Scientific Computing Center at INFN-PISA.  It is a pleasure to
thank A.~Pelissetto and E.~Vicari for discussions and comments.

\end{document}